\documentclass[fleqn,usenatbib]{mnras}

\usepackage[T1]{fontenc}

\DeclareRobustCommand{\VAN}[3]{#2}
\let\VANthebibliography\thebibliography
\def\thebibliography{\DeclareRobustCommand{\VAN}[3]{##3}\VANthebibliography}

\usepackage{graphicx}	
\usepackage{amsmath}	
\usepackage{amssymb}	
\usepackage{float}
\usepackage{natbib}
\usepackage{caption}
\usepackage{color,soul}
\usepackage{gensymb}
\usepackage[utf8]{inputenc}

\usepackage{soul}

\usepackage[usenames,dvipsnames]{xcolor}

\usepackage{newtxtext,newtxmath}

\title[Galaxy mergers in {\tt GALFORM}]{Statistics of galaxy mergers: bridging the gap between theory and observation}

\author[F. Huško, C. G. Lacey \& C. M. Baugh]{
Filip Huško,$^{1}$\thanks{E-mail: filip.husko@durham.ac.uk}
Cedric G. Lacey,$^{1}$
Carlton M. Baugh$^{1}$
\\
$^{1}$ Institute for Computational Cosmology, Department of Physics, University of Durham, South Road, Durham, DH1 3LE, UK\\
}

\date{Accepted XXX. Received YYY; in original form ZZZ}

\pubyear{2020}

\begin{document}

\label{firstpage}
\pagerange{\pageref{firstpage}--\pageref{lastpage}}
\maketitle

\begin{abstract}
We present a study of galaxy mergers up to $z=10$ using the Planck Millennium cosmological dark matter simulation and the {\tt GALFORM} semi-analytical model of galaxy formation. Utilising the full ($800$ Mpc)$^3$ volume of the simulation, we studied the statistics of galaxy mergers in terms of merger rates and close pair fractions. We predict that merger rates begin to drop rapidly for high-mass galaxies ($M_*>10^{11.3}-10^{10.5}$ $M_\odot$ for $z=0-4$), as a result of the exponential decline in the galaxy stellar mass function. The predicted merger rates for massive galaxies ($M_*>10^{10}$ M$_\odot$) increase and then turn over with increasing redshift, by $z=3.5$, in disagreement with hydrodynamical simulations and semi-empirical models. In agreement with most other models and observations, we find that close pair fractions flatten or turn over at some redshift (dependent on the mass selection). We conduct an extensive comparison of close pair fractions, and highlight inconsistencies among models, but also between different observations. We provide a fitting formula for the major merger timescale for close galaxy pairs, in which the slope of the stellar mass dependence is redshift dependent. This is in disagreement with previous theoretical results that implied a constant slope. Instead we find a weak redshift dependence only for massive galaxies ($M_*>10^{10}$ M$_\odot$): in this case the merger timescale varies approximately as $M_*^{-0.55}$. We find that close pair fractions and merger timescales depend on the maximum projected separation as $r_\mathrm{max}^{1.32}$, in agreement with observations of small-scale clustering of galaxies. 
\end{abstract}





\section{Introduction}

Early observations and theoretical considerations showed that {\it in situ} star formation is an ongoing process by which galaxies increase their stellar mass (e.g. Schmidt\citealt{Schmidt}, Kennicutt\citealt{Kennicutt}, Gallego et~al. \citealt{Gallego}). Some galaxies are observed in various stages of close dynamical interaction suggesting an imminent merger (e.g. Toomre \& Toomre\citealt{Toomre}). Mergers can trigger further in situ star formation and bring in {\it ex situ} stellar mass that formed earlier in progenitor galaxies. In situ star formation in galaxies dominates over the mass brought in and reassembled by mergers, according to observations and successful modelling (e.g. Robotham et~al. \citealt{Robotham}, Rodriguez-Gomez et~al. \citealt{RodrGomez2016}, Qu et~al. \citealt{Qu}). Nevertheless, mergers occur between all types of galaxies at all cosmic epochs: the only variation is in their frequency (e.g. Amorisco et~al. \citealt{Amorisco}). 

Mergers have many secondary effects on the properties of galaxies. They are the primary drivers of the transformation of disk galaxies into massive ellipticals (e.g. Toomre \& Toomre\citealt{Toomre}, Schweizer\citealt{Schweizer}, Barnes\citealt{Barnes}, Barnes \& Hernquist\citealt{Barnes1992}, Mihos\citealt{Mihos95}). Mergers trigger bursts of star formation (Schweizer\citealt{Schweizer87}, Barnes \& Hernquist\citealt{Barnes1991}, Mihos \& Hernquist\citealt{Mihos}, Luo et~al. \citealt{Luo}), change the overall distribution and kinematics of stars (Mihos\citealt{Mihos95}, Naab et~al. \citealt{Naab}, Ferreras et~al. \citealt{Ferreras}) and contribute to the growth of supermassive black holes in galactic centres, by facilitating both gas accretion and black hole mergers (Volonteri et~al. \citealt{Volonteri}, Dotti et~al. \citealt{Dotti}, Treister et~al. \citealt{Treister}, Rosario et~al. \citealt{Rosario}, Ellison et~al. \citealt{Ellison}).

The most important effect of mergers is on the evolution of the stellar mass of galaxies. Mergers provide an additional channel for mass growth alongside in situ star formation. Their impact can be quantified through the mass growth rate, $\mathrm{d}M_*/\mathrm{d}t$ (e.~g. Moster et~al. \citealt{Moster2013}, Rodriguez-Gomez et~al. \citealt{RodrGomez2016}). The mass growth rate can be compared directly with the star formation rate, to see when one dominates over the other. A closely related quantity is the \textit{ex situ} fraction, which is the fraction of stellar mass accreted in mergers compared with that formed in situ. Recent theoretical studies have yielded qualitatively similar results for the local Universe (Rodriguez-Gomez et~al. \citealt{RodrGomez2016}, Dubois et~al. \citealt{Dubois}, Henriques et~al. \citealt{Henriques}, Lee \& Yi\citealt{LeeYi}). These studies suggest that the contribution of mergers to the growth of stellar mass increases rapidly for galaxies with mass $> 10^{10.5}-10^{11}$ M$_\odot$. While the ex situ fraction is an interesting quantity, it is impossible to trace the mass evolution of individual galaxies in the real Universe. Observational studies focus instead on the frequency of mergers (e.g. Bundy et~al. \citealt{Bundy}, Robotham et~al. \citealt{Robotham}, Mundy et~al. \citealt{Mundy}). The growth of stellar mass due to mergers can then be inferred from measured merger rates. In addition, observational measurements provide an important check on theoretical models and their predictions.

In general two approaches are used to quantify the merger rate: the merger rate \textit{per galaxy} $\mathrm{d}N/\mathrm{d}t$ and the merger rate \textit{density}: $\mathrm{d}^3n/\mathrm{d}\log M \, \mathrm{d}\,V\mathrm{d}t$. The latter measures the number density of mergers in time, space and mass, which means that it is directly dependent on the number densities of both the primary and secondary galaxies participating in mergers. The merger rate per galaxy, on the other hand, depends on the number density of the secondary galaxies only. 

The merger rate can be theoretically predicted from semi-analytical models of galaxy formation (e.g. Guo \& White\citealt{GuoWhite}, Kitzbichler \& White\citealt{KitzWhite}, hereafter KW08), semi-empirical models (e.g. Stewart et~al. \citealt{Stewart}, Hopkins et~al. \citealt{Hopkins}) and hydrodynamical simulations (e.g. Rodriguez-Gomez et~al. \citealt{RodrGomez2016}, Lagos et~al. \citealt{Lagos}). It can also be estimated from observational data (e.g. Xu et~al. \citealt{Xu}, Mundy et~al. \citealt{Mundy}), although this relies on theoretical assumptions about merger timescales. The easiest way to predict the merger rate is to use semi-analytical models run on outputs of dark matter simulations, since the backbone of these models is the construction of halo and galaxy merger trees. Mergers can be identified by connecting galaxies between the outputs of the simulation, and merger rates calculated through a division by the time interval between two outputs. 

Unfortunately, it is impossible to directly measure merger rates observationally. An often used proxy is the close pair fraction of galaxies, which is divided by an assumed merger timescale to give a merger rate.  However, these timescales remain uncertain to a factor of $2$-$3$ (Conselice\citealt{Conselice2006}, Lotz et~al. \citealt{Lotz2008}, KW08, Conselice\citealt{Conselice2009}, Lotz et~al. \citealt{Lotz2010a}, Lotz et~al. \citealt{Lotz2010b}, Hopkins et~al. \citealt{Hopkins2010b}). Furthermore, the timescales used are usually derived by comparison of the pair fractions and merger rates predicted by a theoretical model. Hence, any application of a merger timescale to an observed pair fraction will yield a merger rate which may be biased towards that model.

Observational studies of merger rates are also marked by inconsistencies in selection criteria. There is some disagreement between the definition of major mergers, which are usually defined as those in which the galaxy pair has a ratio of quantities,  $\mu$, larger than some threshold value. The quantity could be stellar mass, luminosity or flux, and the typical values for this threshold are 1/2.5, 1/3, 1/4 or 1/6. After choosing galaxy pairs according to a chosen threshold, some studies also employ selections that are designed to remove pairs which are not likely to merge (e.g. Lotz et~al. \citealt{Lotz2011}, Casteels et~al. \citealt{Casteels}). This is generally done by studying the morphologies of the galaxies and discarding those that, for example, are not asymmetrical enough to suggest a dynamical interaction. If such a selection is used then the conversion to a merger rate also requires the use of a different merger timescale.

In order to calculate the close pair fraction, observational studies employ selection criteria whereby galaxies are considered to be paired only if they fall within a certain projected distance and a maximum velocity separation. There is disagreement between the values chosen from study to study. However, close pair fractions can be converted from one selection to another (with differing maximum projected separations) under the assumption that the dependence on this quantity is a power law, as suggested by studies of galaxy clustering (e.~g. Le Fèvre et~al. \citealt{LeFevre}, Zehavi et~al. \citealt{Zehavi}). However, these clustering studies are usually restricted to separations of $r>100$ $h^{-1}$kpc, which is well outside the typical maximal separations adopted for close pair studies ($\approx20$ $h^{-1}$kpc). The results of KW08 imply a linear dependence of the close pair fraction on maximal separation. The validity of this assumption has not been tested in detail, nor has the dependence of the close pair fraction (and consequently the merger timescale) on the maximum velocity separation.

Observed close pair fractions can differ significantly, even when common selection criteria are used, due to differing methodologies. Recent observations of the dependence of close pair fractions on redshift do not show convergence between different studies (Man et~al. \citealt{Man}, Ventou et~al. \citealt{Ventou2017}, Mundy et~al. \citealt{Mundy}, Mantha et~al. \citealt{Mantha}, Ventou et~al. \citealt{Ventou2019}, Duncan et~al. \citealt{Duncan}). Most of these measure a close pair fraction that plateaus or decreases above some redshift (with the exception of Duncan et~al. \citealt{Duncan}, who measure a rising close pair fraction out to $z=6$). However, the details differ significantly, with Ventou et~al.~\cite{Ventou2017}, Mantha et~al.~\cite{Mantha} and Ventou et~al.~\cite{Ventou2019} measuring pair fractions that decrease sharply (by $z=2-3$, depending on the mass selection), while Man et~al.~\cite{Man} and Mundy et~al.~\cite{Mundy} find pair fractions that plateau or decrease only slightly above some redshift. Even these two studies, which find a similar functional dependence of the pair fraction on redshift, disagree on the normalisation. Given these differences, it is interesting to see which of these studies (if any) agree with  theoretical models, particularly at high redshifts.

Here, we present a detailed study of galaxy merger statistics using the {\tt GALFORM} semi-analytical model of galaxy formation (e.g. Cole et~al. \citealt{Cole}, Bower et~al. \citealt{Bower}, Gonzalez-Perez et~al. \citealt{GonzPerez}, Lacey et~al. \citealt{Lacey}, Baugh et~al. \citealt{Baugh}), which is run on outputs of the Planck Millennium dark matter simulation. Our aim is to study theoretical galaxy merger rates and close pair fractions of galaxies with unprecedented precision and determine the stellar mass dependencies of these quantities in detail. We also aim to determine the redshift dependencies of these quantities up to previously unprobed redshifts, providing predictions for upcoming high-redshift observatories (e.g. \textit{JWST}). This is possible as we utilise the full volume of the Planck Millennium simulation: ($800$ Mpc)$^3$.

Once merger rates and close pair fractions are calculated, the dependence of the merger timescale on stellar mass and redshift follows. If pairs are chosen with a variety of close pair selection criteria, the dependence of the close pair fraction (and merger timescale) on the maximum projected separation and velocity separation can also be determined. The merger timescale calculated in this way can be used to obtain merger rates from close pair fractions made with an arbitrary selection. Furthermore, the dependence on maximal projected separation and line-of-sight velocity can be used to convert a close pair fraction from one selection to another, enabling a consistent comparison between different studies.

The outline of this paper is as follows. In \S~\ref{sec:Intro} we present the N-body simulation in which we set the semi-analytical model and discuss the treatment of galaxy mergers. \S~\ref{sec:Method} describes our methods of calculating merger rates, close pair fractions and merger timescales. We also discuss some observational studies, their differences and ways of converting their results from one selection to another. In \S~\ref{sec:MergRates} we present our results for merger rates as functions of stellar mass and redshift. We also compare in detail our merger rates with observations. \S~\ref{sec:CPF} presents our results on the close pair fraction and its dependence on stellar mass and redshift, as well as a comparison with observations. In \S~\ref{sec:TSres} we present our results for the effective merger timescale of close pairs and its dependence on stellar mass and redshift, as well as on projected and velocity separation, and derive fitting formulae for these dependencies. In \S~\ref{sec:sum} we summarise and conclude.




\section{N-body simulation and {\tt GALFORM}}
\label{sec:Intro}
\subsection{Galaxy formation model}

We use the {\tt GALFORM} semi-analytical model of galaxy formation implemented in the Planck Millennium N-body simulation of the evolution of structure in the dark matter (Baugh et~al. \citealt{Baugh}). {\tt GALFORM} models various physical processes, such as dark matter halo assembly, shock heating and radiative cooling of gas, the formation of galaxy disks, ejection and heating of gas due to supernova and AGN feedback, galaxy mergers and disk instabilities, as well as their effects on galaxy mass and morphology, chemical evolution of the gas and stars, the stellar luminosity of galaxies, and dust emission/absorption (Cole et~al. \citealt{Cole}; Baugh \citealt{Baugh2006}; Bower et~al. \citealt{Bower}; Lacey et~al. \citealt{Lacey}). 

The Planck Millennium N-body simulation (Baugh et~al. \citealt{Baugh}) uses the cosmological parameters inferred from the first year \citealt{PLANCK} data release\footnote{The cosmological parameters used are: $\Omega_\mathrm{M}=0.307$, $\Omega_\mathrm{\Lambda}=0.693$, $\Omega_\mathrm{b}=0.0483$, $h=0.677$, $\sigma_{8} = 0.8288$ and $n_{\rm s}=0.9611$.}. The simulation has a volume of (800 Mpc)$^3$ and uses 5040$^{3}$ particles. The minimum halo mass is set to 20 particles, corresponding to 2.12$\times10^9$ $h^{-1}\mathrm{M}_\odot$. The halo merger trees are stored at 269 output times.  
Halo merger trees are constructed using the {\tt SUBFIND} halo finder and the {\tt DHALOS} algorithm  described in Jiang et~al.~\cite{Jiang2014}. 

The volume of the Planck Millennium simulation, the large number of outputs and the mass resolution allow us to make accurate predictions of galaxy merger rates, despite galaxy mergers being relatively rare events. We are able to produce predictions for merger rates and close pair fractions in 40 bins in stellar mass between $10^8$ M$_\odot$ and $10^{12}$ M$_\odot$, as well as 40 redshift bins between $z=0$ and $z=10$. The simulation volume is $\approx500$ times larger than the original Illustris and EAGLE simulation boxes (Vogelsberger et~al. \citealt{Vogelsberger}, Schaye et~al. \citealt{Schaye}), allowing the merger statistics of high mass galaxies ($M_* > 10^{11}$ M$_\odot$) to be studied with high precision for the first time.  

The {\tt GALFORM} model used in this analysis is the one from Lacey et~al.~\cite{Lacey}, with the small recalibration of parameters made in Baugh et~al.~\cite{Baugh} for the Planck Millennium cosmology and an updated galaxy merger scheme (see \S\ref{sec:galaxy_merger_scheme}).  As shown in Lacey et~al.~\cite{Lacey}, {\tt GALFORM} successfully reproduces the optical and near-IR luminosity functions, the fractions of early type galaxies, and the Tully-Fisher, metallicity-luminosity and size-luminosity relations at $z=0$, as well as far-IR and sub-mm number counts, and far-UV luminosity functions between $z=3$ and $z=6$. The HI mass function and HI mass -  halo mass relation are studied in detail in Baugh et~al.~\cite{Baugh}, where they are shown to agree well with observations.

\begin{figure}
\includegraphics[width=0.99\columnwidth, trim = 0 15 0 0]{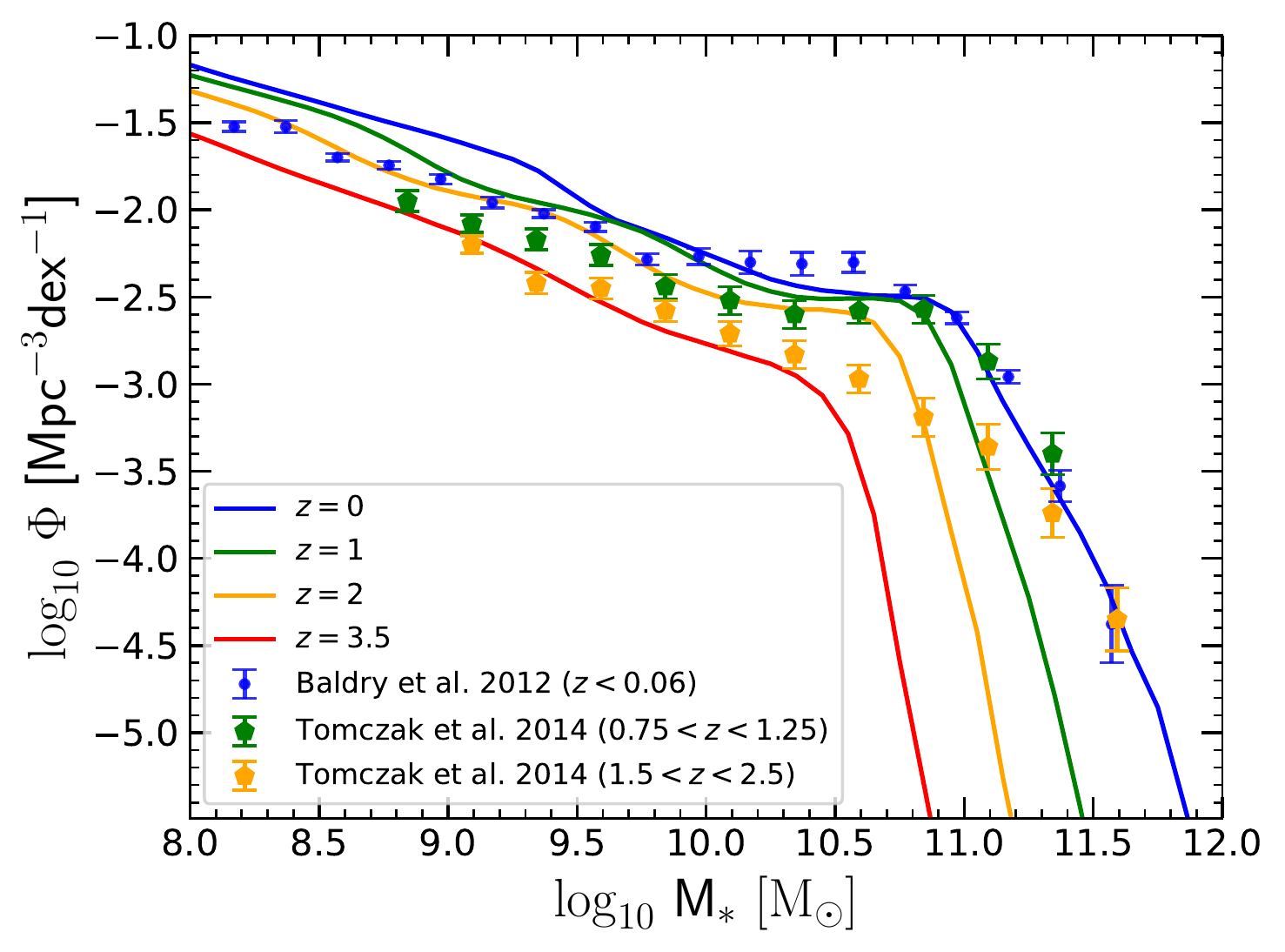}
\caption{The galaxy stellar mass function predicted by {\tt GALFORM} at several different redshifts, as indicated by the legend. Dots and error bars of different colours (corresponding to lines) represent observational estimates of the GSMF from Baldry et~al. \protect\cite{Baldry} and Tomczak et~al. \protect\cite{Tomczak}.}
\label{fig:figSMF}
\end{figure}

The galaxy stellar mass function (GSMF hereafter) predicted by {\tt GALFORM} is shown at several redshifts in Fig.~\ref{fig:figSMF}, along with  observational measurements. At higher redshifts we show measurements only from Tomczak et al.~\cite{Tomczak}, but these are consistent with other works (e.g. Ilbert et al.\citealt{Ilbert}, Muzzin et al. \citealt{Muzzin}, Davidzon et al.\citealt{Davidzon}, Wright et al.\citealt{Wright}, McLeod et al.\citealt{McLeod}). The predicted GSMF agrees only roughly with the observational estimates, particularly below the break. 
As noted in Mitchell et~al.~\cite{Mitchell} (see also Gonzalez-Perez et~al.~\cite{GonzPerez}, and Lacey et~al.~\cite{Lacey}), the comparison with the observed GSMF should not be made using the predicted stellar masses directly. This is because observational GSMFs are inferred by fitting model SEDs to multi-band observed fluxes. Mimicking this procedure in the model leads to much better agreement (e.g. Fig.~A7 in Gonzalez-Perez et~al. \citealt{GonzPerez}, Fig.~24 in Lacey et~al.\citealt{Lacey}). At higher redshifts, however, we find that {\tt GALFORM} predicts too few massive galaxies. The disagreement is large enough that SED fitting does not help to bring {\tt GALFORM} fully in line with observed data for $z>3$.

\subsection{Galaxy mergers}
\label{sec:galaxy_merger_scheme}
In {\tt GALFORM}, mergers are assumed to occur only between a satellite galaxy and a central galaxy in the same host dark matter halo, after dynamical friction has caused the orbit of the satellite to decay. 
In this paper we use the improved treatment of galaxy mergers introduced into {\tt GALFORM} by Simha \& Cole\cite{Simha} and used in Baugh et~al.\citet{Baugh}. In earlier versions of {\tt GALFORM}, for example in Lacey et~al.~\cite{Lacey}, haloes were tracked only up to when they entered a more massive halo and became a satellite halo. At this point, the timescale for the satellite galaxy to merge with the central galaxy was calculated using an analytical formula. {\tt GALFORM} originally used the merger timescale formula from Lacey \& Cole\cite{LaceyCole}, which was derived by integrating the Chandrasekhar\cite{Chandrasekhar} dynamical friction rate along orbits in a singular isothermal sphere halo, ignoring tidal stripping of the satellite halo. In Lacey et~al.~\cite{Lacey}, this was replaced by the formula from Jiang et~al.~\cite{Jiang}, which has a similar general form, but was calibrated to hydrodynamical simulations of galaxy formation, and so included tidal stripping effects. In the new {\tt GALFORM} merger scheme (Simha \& Cole\cite{Simha}), the subhaloes containing galaxies continue to be tracked in the dark matter simulation after they become satellites. When the satellite halo can no longer be resolved in the simulation, the remaining time $\tau_{\rm mg}$ for the satellite galaxy to merge with the central galaxy is calculated using an analytical formula which is a modified version of that in Lacey \& Cole\cite{LaceyCole}, but applied at the last point at which the satellite halo was identified in the dark matter N-body simulation.

The Simha \& Cole\cite{Simha} formula for the remaining time until the galaxy merges is  
\begin{equation}
\begin{aligned}
\tau_\mathrm{mg}= & \frac{\epsilon^\beta}{0.86}\frac{M_\mathrm{pri}(<r)}{M_\mathrm{sat}}\frac{1}{\ln(1+M_\mathrm{pri}(<r)/M_\mathrm{sat})}\bigg(\frac{r_\mathrm{c}}{r}\bigg)^\alpha T_\mathrm{dyn},
\end{aligned}
\label{eq:TmergSimha}
\end{equation}
in which all quantities are calculated at the last timestep at which the satellite halo was resolved in the N-body simulation. In the above, $r$ is the distance of the satellite from the centre of the main halo, $M_\mathrm{sat}$ is the mass of the satellite halo, $M_\mathrm{pri}(<r)$ is the mass of the main halo within radius $r$ and $T_\mathrm{dyn} = r/V_c(r)$ is the  dynamical timescale at radius $r$. $\epsilon$ is the circularity of the satellite orbit, defined as the ratio of the angular momentum to that of a circular orbit with the same energy and corresponding radius $r_\mathrm{c}$. The best values for the exponents $\alpha$ and $\beta$ were found by Simha \& Cole by requiring consistency between the numbers of satellite haloes found in the Millennium~I (Springel et~al. \citealt{Springel}) and Millennium~II (Boylan-Kolchin et~al. \citealt{Boylan2008}) simulations, which have very different mass resolutions. This gave $\alpha=1.8$ and $\beta=0.85$.


The formula given by Eqn.~(\ref{eq:TmergSimha}) has a similar form to that in Lacey \& Cole\cite{LaceyCole}. Lacey \& Cole used a different Coulomb logarithm given by $\ln \Lambda=\ln(M_\mathrm{pri}/M_\mathrm{sat})$ instead of $\ln \Lambda=\ln(1+M_\mathrm{pri}/M_\mathrm{sat})$, and found $\alpha=2$ and $\beta=0.78$. 
In \S~\ref{sec:TSres} we  show that use of Eqn.~(\ref{eq:TmergSimha}) leads to small-scale clustering of galaxies which is in agreement with observations.





\section{Methods}
\label{sec:Method}

Here, we describe how we  calculate galaxy merger rates and close pair fractions. We discuss the selection criteria used for close pairs in the context of observational studies, and describe how to compare consistently between studies with different selection criteria.

\subsection{Merger rates}
\label{sec:MRmethod}

We use the galaxy merger trees constructed by {\tt GALFORM} to calculate merger rates. Each galaxy is assigned a unique number (ID) at each snapshot. For a given snapshot, we list all galaxies according to their descendant IDs at a later snapshot. Galaxies with the same descendant ID are identified as being about to merge. We assume that all galaxies with the same descendant ID merge with the most massive progenitor, and bin all pairs by its mass. The merger rate per galaxy is
\begin{equation}
\frac{\mathrm{d}N}{\mathrm{d}t}=\frac{\Delta N_\mathrm{merg}}{\Delta N \Delta t},
\label{eq:eqMergRatePerGal}
\end{equation}
where $\Delta N_\mathrm{merg}$ is the total number of pairs in the mass bin set by the most massive progenitor, and satisfying a stellar mass ratio condition ($\mu_* \in [0.1,0.25]$ for minor mergers and $\mu_* \in [0.25,1]$ for major mergers). $\Delta N$ is the number of galaxies in the mass bin and $\Delta t$ is the time interval between two snapshots. The merger rate density is 
\begin{equation}
\frac{\mathrm{d}^3 n}{\mathrm{d}t\mathrm{d}V\mathrm{d}\log M_*}=\frac{\Delta N_\mathrm{merg}}{\Delta t \Delta V \Delta \log M_*},
\label{eq:eqMergRateDen}
\end{equation}
where $\Delta \log M_*$ is the logarithmic mass bin width and $\Delta V$ the simulation volume.

The assumption that all galaxies merge with the most massive progenitor, as opposed to sometimes merging with each other, is reasonable. The large number of snapshots we have available (269) helps minimise any errors due to sequential galaxy mergers.

\subsection{Close pair fraction}
\label{sec:CPmethod}

In observational studies, two methods are generally used to  select galaxies as candidates for merger pairs. The close pair method  (e.g. Xu et~al. \citealt{Xu}, Robotham et~al. \citealt{Robotham}, Mundy et~al. \citealt{Mundy}) imposes certain selection criteria (usually a maximum projected separation, maximum velocity separation and minimum mass ratio). The other method (e.g. Lotz et~al. \citealt{Lotz2008}, Casteels et~al. \citealt{Casteels}) in addition tries to filter out pairs that are not going to merge, by requiring galaxies to display asymmetry. 
We extract pair fractions from our simulation using the first method. While this has the disadvantage of including galaxies that are not physically associated, this can be taken into account when calculating merger rates by using an appropriate merger timescale (see KW08 and \S \ref{sec:ObsTimeScale} and \ref{sec:TSres}).

Rather than constructing galaxy light cones to calculate close pair fractions of galaxies, as done by e.g. KW08 and Snyder et~al. \citealt{Snyder}, we use a simpler, albeit slightly more approximate, method which yields much larger samples of close pairs. We choose a fixed axis as the line of sight (e.g. the $z-$axis), while the other two axes are used for calculating projected separations. This allows us to use the full simulation volume at every snapshot. 

Our calculation of close pair fractions is complicated by the existence of satellite galaxies whose subhaloes are no longer resolved in the simulation due to tidal stripping. 
In the improved {\tt GALFORM} merger scheme, satellite galaxies are only able to merge once their host subhaloes can no longer be resolved. In this case, we assume that the current orbital radius, $r$, of such subhaloes can be calculated using:
\begin{equation}
\bigg(\frac{r}{R}\bigg)^{\alpha}=\frac{T_\mathrm{merg,remaining}}{T_\mathrm{merg}},
\label{eq:RScale}
\end{equation}
where $R$ is the initial separation of the subhalo and primary when the subhalo is `lost' and $T_\mathrm{merg}$ is the merger time calculated at that point.  $T_\mathrm{merg,remaining}=T_\mathrm{merg}-t_\mathrm{elapsed}$ is the remaining time until the merger. This relation assumes that the radial decay of orbits satisfies the merger timescale given by Simha \& Cole\cite{Simha} (Eqn.~\ref{eq:TmergSimha}). The velocities of such subhaloes are kept the same as when they were last resolved. 

We consider a galaxy to have a major close pair if there is another galaxy of similar mass ($\mu_*>0.25$) within a projected distance $r_\mathrm{sep}<r_\mathrm{max}$ (using the $x$ and $y$ coordinates in the simulation), and with velocity separation (along the $z$ axis, including the Hubble flow)  $v_{z,\mathrm{sep}}<v_{z,\mathrm{max}}$. The major close pair fraction with arbitrary selection criteria $r_\mathrm{max}$ and $v_\mathrm{max}$ then follows as
\begin{equation}
f_\mathrm{maj}=\frac{\Delta N_\mathrm{maj}}{\Delta N},
\label{eq:eqClosePairFrac}
\end{equation}
where $\Delta N_\mathrm{maj}$ is the number of close pairs in a given mass bin, while $\Delta N$ is the total number of galaxies in that mass bin.

\subsection{Obtaining merger rates from close pair fractions}
\label{sec:ObsTimeScale}

Merger rates cannot be measured directly from observations. They are usually inferred by assuming a relation between the close pair or merger fraction and merger rate:
\begin{equation}
\frac{\mathrm{d}N}{\mathrm{d}t}=\frac{f}{T_\mathrm{mg}},
\label{eq:eqCPTS}
\end{equation}where $T_\mathrm{mg}$ is the effective merger timescale. Note that $T_\mathrm{mg}$ is a different quantity from $\tau_\mathrm{mg}$ (defined in Eqn.~\ref{eq:TmergSimha}), which is used to calculate actual merger times in the simulation based on subhalo positions. $T_\mathrm{mg}$, on the other hand, should be viewed mostly as a mathematical construction, whose purpose is to convert a close pair fraction or a merger fraction to a merger rate. If one uses the close pair fraction, then the corresponding merger timescale may have little physical meaning in some regimes (e.g. for lower masses, where one would expect large numbers of spurious pairs; see \S~\ref{sec:CPF}). 

In the close pair method, the effective merger timescale can be calculated by relating merger rates to close pair fractions in simulations (e.g. KW08, Snyder et~al. \citealt{Snyder}). In the merger fraction method, the merger timescale can be obtained by considering how long merging galaxies appear as a pair similar to those in observations (e.g. Lotz et~al. \citealt{Lotz2010a}). Merger timescales are still uncertain to a factor of 2-3, and are thus the largest source of uncertainty in observational merger rates.

The differences between merger timescale estimates can be traced to their definition, which is to convert a \textit{close} pair fraction or a \textit{merger} fraction into a merger rate. Close pair galaxies are chosen solely by dynamical selection criteria (proximity in projected space and line-of-sight velocity). Merger fractions, on the other hand, are calculated based on the morphology of pairs of galaxies and by attempting to decide if they represent a merging system. The merger fraction method was very popular in earlier observational studies (Patton et~al. \citealt{Patton}, de Propris et~al. \citealt{dePropris}, Ryan et~al. \citealt{Ryan}, Lotz et~al. \citealt{Lotz2008a}, Conselice et~al. \citealt{Conselice2008} $-$\citealt{Conselice2009}, Lopez-Sanjuan et~al. \citealt{LopezSanjuan2011};\citealt{LopezSanjuan}, Bluck et~al. \citealt{Bluck2012}, Stott et~al. \citealt{Stott}). However, the close pair method has been the focus of most recent studies (e.g. Man et~al. \citealt{Man}, Mundy et~al. \citealt{Mundy}, Ventou et~al. \citealt{Ventou2017}, Mantha et~al. \citealt{Mantha}, Duncan et~al. \citealt{Duncan}, Ventou et~al. \citealt{Ventou2019}). For this reason, and because close pair fractions are generally easier to calculate from theoretical models, we focus on this method.

One of the goals in this paper is to compare our predictions to different models or observations in a consistent manner. Since most observational studies use merger timescales based on KW08, we do the same when calculating observational merger rates.
The KW08 merger timescale takes the form
\begin{equation}
T_\mathrm{mg,KW}=2 \hspace{0.5mm}\mathrm{Gyr} \left( \frac{r_\mathrm{max}}{50\hspace{0.5mm} \mathrm{kpc}} \right)
\bigg( \frac{M_*}{10^{10.6}\hspace{0.3mm} h^{-1}\mathrm{M}_\odot}\bigg)^{-0.3} \bigg(1+\frac{z}{8} \bigg),
\label{eq:KW}
\end{equation}
where $r_\mathrm{max}$ is the maximal projected separation of pairs. However, KW08 also offer a more accurate formula, which works better for lower-mass galaxies. We do not reproduce it here due to its complexity (see Table 1 in KW08 for further details). The fit in question is valid for pairs with a maximal projected separation $r_\mathrm{max}=30$ $h^{-1}$kpc and velocity separation $v_\mathrm{max}=300$ kms$^{-1}$. If an observational study originally used KW08 merger timescales, we compare merger rates directly. However, if a different merger timescale was used, we recalculate the merger rates using the KW08 relations, and use merger rates calculated in such a way as a basis of comparison. Finally, using the KW08 formula allows us to calculate merger rates using measurements from observational studies which only provide close pair fractions. In Table~\ref{tab:tab0}, we provide details for all close pair studies that closely match our definition (see next subsection). This table also specifies for each study whether merger rates were calculated using the KW08 formulae.

\subsection{A standard selection of close pairs}
\label{sec:ObsDiscr}

\captionsetup[table]{skip=0pt} 
\begin{table*}
\begin{center}
\caption{A summary of close pair studies. $^{*}$Threshold values of $\log_{10}M_*$ used by the study, in units of M$_\odot$. Where multiple values are cited, these represent different datasets against which we compare. $^\dagger$Our standard definition is in terms of $\mu_*$, the minimal value for a merger to be considered major. $^\ddagger$Where in units of kms$^{-1}$, the study uses spectroscopic redshifts, and this value gives the maximal velocity separation. Where unitless, photometric redshifts were used, with the value referring to the redshift separation $\Delta z$. In this case, $z_1$ and $z_2$ are the redshifts of the primary and secondary galaxy, respectively, while $z_{1,2}$ is their mean redshift. $\sigma_{z_{1,2}}$ are the associated uncertainties. For photometric studies, the velocity criterion is always significantly greater than $v_\mathrm{max}=1000$, so we assume that the number of pairs has saturated (see \S~\ref{sec:TSrv}). $^\#$ Whether a study originally calculated merger rates using KW08 merger timescales (e.g. Eqn.~\ref{eq:KW}). If not, we used their close pair fractions to calculate merger rates using KW08. $^\Delta$The conversion factor which brings the described selection (of a given study), to the selection $\mu_*=1/4$, $r_\mathrm{sep}<20$ $h^{-1}$kpc and $v_\mathrm{sep}<500$ kms$^{-1}$, i.e. our standard selection of close pairs (see text for details). These conversion factors were calculated using Eqn.~(\ref{eq:conv}).}
\label{tab:tab0}
\end{center}
\begin{tabular*}{1.01\textwidth}{c}
\hspace{63.5mm}OBSERVATIONAL CLOSE PAIR STUDIES 
\end{tabular*}\\

\begin{tabular*}{1.01\textwidth}{@{\extracolsep{\fill}}lccccccc}
  \hline \hline
  Study  & Mass range$^*$ & Redshift range & Major mergers$^\dagger$ & $r_\mathrm{sep}\hspace{0.5mm}[\mathrm{kpc}]$ & $v_\mathrm{sep}^\ddagger$ & KW08$^\#$ & Conversion factor$^\Delta$ \\
  \hline \hline
  Bundy (2009) & $>10,10.5-11^{\mathrm{b}_3},>11$ & $[0.4,1.4]$ & $\Delta M_\mathrm{K}<1.5^{\mathrm{c}}$ & $[5,20]h^{-1}$ & $ <(\sigma_{z_1}^2+\sigma_{z_2}^2)^{1/2,\mathrm{d}_1}$ & N & 0.812 \\
  \hline
  Domingue (2009) & $[9,12]$ & $[0.034,0.12]$ & $\mu_*=1/2.5$ & $[5,20]h^{-1}$ & $ <1000\hspace{0.5mm}\mathrm{kms}^{-1}$ & Y & 1.576 \\
  \hline
  Robaina (2010) & $>10.7^{\mathrm{b}_4}$ & $[0.2,1.2]$ & $\mu_*=1/4$ & $<30$ & $<\sqrt{2}\sigma_\mathrm{z_1}$ & N & 0.760 \\
  \hline
  Williams (2011) & $>10.8^{\mathrm{b}_4}$ & $[0.4,2]$ & $\mu_*=1/4$ & $[10,30]h^{-1}$ & $<0.2(1+z_{1,2})$ & Y & 0.607 \\
  \hline
  LS (2012)$^{\mathrm{a_1}}$ & $>11$ & $[0.2,0.9]$ & $\mu_*=1/4$ & $[10,30]h^{-1}$ & $<500\hspace{0.5mm}\mathrm{kms}^{-1}$ & N & 0.748 \\
  \hline
  Newman (2012) & $>10.7^{\mathrm{b}_4}$ & $<2.5$ & $\mu_*=1/4$ & $[10,30]h^{-1}$ & $<z_0(1+z_{1,2})^{\mathrm{d}_2}$ & N & 0.607 \\
  \hline
  Xu (2012) & $[9.4,11.6]$ & $<1$ & $\mu_*=1/2.5$ & $[5,20]h^{-1}$ & $<500\hspace{0.5mm}\mathrm{kms}^{-1}$ & Y$^{\mathrm{e}_1}$ & 1.763 \\
  \hline
  Robotham (2014) & $[8,12]$ & $[0.01,0.2]$ & $\mu_*=1/3$ & $<20h^{-1}$ & $<500\hspace{0.5mm}\mathrm{kms}^{-1}$ & Y & 1.309 \\
  \hline
  Man (2016) & $>10.8^{\mathrm{b}_4}$ & $[0.1,3]$ & $\mu_*=1/4$ & $[10,30]h^{-1}$ & $<z_0(1+z_{1,2})^{\mathrm{d}_2}$ & N & 0.607 \\
  \hline
  Mundy (2017) & $>10,11$ & $<3.5$ & $\mu_*=1/4$ & $[5,30]$ & $<\mathrm{CDF}^{\mathrm{d}_3}$ & N & 0.835 \\
  \hline
  Mantha (2018) & $>10.3^{\mathrm{b}_{2}}$ & $<3$ & $\mu_*=1/4$ & $[5,50]$ & $<(\sigma_{z_1}^2+\sigma_{z_2}^2)^{1/2,\mathrm{d}_1}$ & N & 0.399 \\
  \hline
  Duncan (2019) & $>10.3^{\mathrm{b}_{2}}$ & $[0.5,6]$ & $\mu_*=1/4$ & $[5,30]$ & $<500\hspace{0.5mm}\mathrm{kms}^{-1}$ & N & 0.399 \\
  \hline
  Ventou (2019) & $>9.5$ & $[0.2,6]$ & $\mu_*=1/6$ & $[5,50]$ & $<300\hspace{0.5mm}\mathrm{kms}^{-1}$ & N & 0.464 \\
  
  \hline \hline
\end{tabular*}

\begin{tabular*}{1.01\textwidth}{c}
\hspace{63.5mm}THEORETICAL CLOSE PAIR STUDIES 
\end{tabular*}\\

\begin{tabular*}{1.01\textwidth}{@{\extracolsep{\fill}}lccccccc}
  \hline \hline
    Study  & Mass range$^*$ & Redshift range & Major mergers$^\dagger$ & $r_\mathrm{sep}\hspace{0.5mm}[\mathrm{kpc}]$ & $v_\mathrm{sep}^\ddagger$ & KW08$^\#$ & Conversion factor$^\Delta$ \\
  \hline \hline
  Snyder (2017)$^{\mathrm{a}_2}$ & $10.5-11^{\mathrm{b}_3}$ & $<4$ & $\mu_*=1/4$ & $[10,50]$ & $<0.02(1+z_1)$ & N$^{\mathrm{e}_2}$ & 0.336 \\
  \hline
  Endsley (2020)$^{\mathrm{a}_3}$ & N/A$^{\mathrm{b}_5}$ & $[4,10]$ & $\mu_*=1/4$ & $[5,25]h^{-1}$ & $ <1000\hspace{0.5mm}\mathrm{kms}^{-1}$ & N$^{\mathrm{e}_2}$ & 0.697 \\
  \hline
  O'Leary (2021a)$^{\mathrm{a}_4}$ & $>9.5,>10,>11$ & $<6$ & $\mu_*=1/4$ & [0,30] & $ <500\hspace{0.5mm}\mathrm{kms}^{-1}$ & N$^{\mathrm{e}_2}$ & 1 \\
  \hline \hline
\end{tabular*}

\begin{tabular*}{1.01\textwidth}{@{\extracolsep{\fill}}l}
$^{\mathrm{a_1}}$Lopez-Sanjuan (2012). $^{\mathrm{a_2}}$Illustris: Vogelsberger et~al.~\cite{Vogelsberger}. $^{\mathrm{a_3}}$UNIVERSE-MACHINE: Behroozi et~al.~\cite{Behroozi}. $^{\mathrm{a_4}}$EMERGE: Moster et~al.~\cite{Moster}. The \\ 
selections we use here for EMERGE data are different than those shown in the original paper, as we were supplied data that matches our selection by the authors. \\

$^{\mathrm{b}_{1,2,3,4}}$We compare results obtained for these mass selections with our following selections: $\log_{10}M_*>9.5,>10,=10.8$ and $>11$, respectively.\\
$^{\mathrm{b}_5}$Authors did not specify mass range.\\

$^{\mathrm{c}}$While this selection deviates from our standard choice, the authors note that it is approximately equivalent to choosing $\mu_*=1/4$. \\

$^{\mathrm{d}_{1}}$This study used both spectroscopic and photometric redshifts. See paper for exact criterion used. $^{\mathrm{d}_2}$ $z_0=0.1$ for $z<1$ and $z_0=0.2$ for $z>1$.
\\$^{\mathrm{d}_3}$This study used conditional probability density functions to determine redshift differences. See study for more details.\\
$^{\mathrm{e}_1}$This study combined KW08 merger timescale results with those of Lotz et~al.\cite{Lotz2011}. We recalculate their merger rates using the original KW08 formula,\\as for all other close pair studies. $^{\mathrm{e}_2}$ We do not calculate merger rates from theoretical close pair fractions, since intrinsic merger rates are provided.
\end{tabular*}

\end{table*}

In the previous subsection, we outlined how we obtain a consistent estimate of the merger rate from observational studies (if one is not provided). However, a comparison between studies is complicated by varying choices of selection criteria. When comparing results obtained with different selections, it is possible to convert close pair fractions to some standard selection using scaling relations. KW08 state that close pair fractions (and therefore merger timescales) scale linearly with $r_\mathrm{max}$, but provide no detailed analyses of this dependence. Furthermore, they did not consider the dependence of the merger timescale on $v_\mathrm{max}$. For this reason, we choose to rescale close pair fractions from other studies to a standard selection using our own results on merger timescales and pair fractions (see Section \ref{sec:TSres}, Eqn.~\ref{eq:TSfitfinv}): $f_\mathrm{mg}\propto$$r_\mathrm{max}^{1.32}\times v_\mathrm{max}^{0.78}$. These relations are valid for $r_\mathrm{max}\in[10,30]$ $h^{-1}$kpc and $v_\mathrm{max}<500$ kms$^{-1}$. Selections with $v_\mathrm{max}>500$ kms$^{-1}$ require more care, due to saturating numbers of pairs. We use our full dependence on $v_\mathrm{max}$, modeled with an error function (Eqn.~\ref{eq:TSfitv}), where needed. Note that the dependence on $r_\mathrm{max}$ is in good agreement with studies of galaxy clustering. The $v_\mathrm{max}$ dependence, including the saturation, agrees with measurements of de Ravel et~al.~\cite{deRavel}.

Some observational papers also employ a lower cutoff $r_\mathrm{min}$ when selecting close pairs, so that $r_\mathrm{sep}\in[r_\mathrm{min}$,$r_\mathrm{max}$]. This is usually done because of sample incompleteness due to source blending. We include the effects of this lower cutoff when rescaling close pair fractions from observational studies, by adding or subtracting the number of pairs expected from our $f_\mathrm{mg}\propto r^{1.32}$ scaling.

Our standard selection when comparing merger rates is $r_\mathrm{min}=0$, $r_\mathrm{max}=30$ $h^{-1}$kpc and $v_\mathrm{max}=300$ kms$^{-1}$, consistent with the merger timescale of KW08 (Eqn.~\ref{eq:KW}), which we use to convert observational close pair fractions into merger rates. When comparing our close pair fractions with observations, we choose a somewhat different selection: $r_\mathrm{max}=20$ $h^{-1}$kpc and $v_\mathrm{max}=500$ kms$^{-1}$. These values are consistent with recent observational studies (e.g. Robotham et~al. \citealt{Robotham}, Mundy et~al. \citealt{Mundy}, Duncan et~al. \citealt{Duncan}). 

Finally, observational results can differ due to different definitions of major and minor mergers. These merger types are usually delimited by a threshold ratio, $\mu$, of quantities, which can be stellar mass or luminosity. As shown in Mantha et~al.~\cite{Mantha}, close pair fractions depend not only on $\mu$ itself, but also on whether it represents a ratio of luminosities or stellar masses. We choose to delimit merger types by the stellar mass ratio.\footnote{Values for the limiting ratio $\mu_*$ are often taken as $1/6$, $1/4$, $1/3$ or $1/2.5$, depending on the study.}. We define major mergers as those with $\mu_*\in[0.25,1]$, while minor mergers satisfy $\mu_*\in[0.1,0.25]$.

Using their observational data, Xu et~al.~\cite{Xu} show that close pair fractions (for major mergers) can be converted from  study A to study B by multiplying the original close pair fraction by $\log_{10}\mu_\mathrm{*,B}/\log_{10}\mu_\mathrm{*,A}$, where $\mu_\mathrm{*,i}$ is the threshold mass ratio for a major merger in study $i$. This relation is consistent with the assumption that the differential number count of close pairs scales as $\mathrm{d}N/\mathrm{d}\mu_* \propto 1/\mu_*$. We find this to be close to our own dependence, $\mathrm{d}N/\mathrm{d}\mu_* \propto 1/\mu_*^{1.25}$, which we find from our own sample of mergers. This dependence is in very good agreement with results from Illustris (Rodriguez-Gomez et~al. \citealt{RodrGomez}). It roughly holds regardless of stellar mass and redshift. The cumulative number of pairs corresponding to a differential distribution of $1/\mu_*^{1.25}$ scales as $\propto(1/\mu_*^{0.25}-1)$. We use this dependence to convert close pair fractions, in case an observational study uses a threshold $\mu_*$ which differs from 0.25. This relation differs somewhat from the logarithmic dependence of Xu et~al.~\cite{Xu}. Note, however, that even their differential number counts are consistent with a slightly steeper dependence on $\mu_*$ than $1/\mu_*$ (see their Fig. 18).

Having discussed the dependence of the close pair fraction on selection criteria $r_\mathrm{max}$, $v_\mathrm{max}$ and $\mu_*$, we can now state our conversion formula. Given a close pair fraction calculated with selection criteria A, we convert the pair fraction to a different selection, B, in the following way:
\begin{equation}
f_\mathrm{B}= \left( \frac{1/\mu_{*,B}^{0.25}-1}{1/\mu_{*,A}^{0.25}-1} \right)
\bigg(\frac{r_\mathrm{max,B}}{r_\mathrm{max,A}}\bigg)^{1.32} \frac{\mathrm{erf}(v_\mathrm{max,B}/V_0)^{0.78}}{\mathrm{erf}(v_\mathrm{max,A}/V_0)^{0.78}} f_\mathrm{A},
\label{eq:conv}
\end{equation}
with $V_0=540$ kms$^{-1}$ a fitting parameter. Note that this is only valid for major mergers (the ones for which we compare our results with other studies). For minor mergers, the $\mu_*$-dependant factor needs to be replaced with an appropriate factor which scales with the total number of pairs between $\mu_{*,1}$ and $\mu_{*,2}$, the lower and upper limit of mass ratios considered as minor mergers, respectively.

In Table~\ref{tab:tab0}, we provide a summary of all studies against which we compare our predicted close pair fractions. Our requirement for a study to be comparable is that it uses stellar masses for sample and pair selections (instead of luminosities), and that it includes all projected pairs. Table~\ref{tab:tab0} also includes conversion factors which convert close pair fractions of that study to our standard selection, described above, using Eqn.~(\ref{eq:conv}).




\section{Merger rates}
\label{sec:MergRates}

Here we present our results on galaxy merger rates. These are calculated from the simulation as described in \S~\ref{sec:MRmethod}. Merger rates from observational studies are converted by applying appropriate conversion factors (as explained in \S~\ref{sec:ObsTimeScale} and \ref{sec:ObsDiscr}) to account for different selection criteria or merger timescales.

\subsection{Dependence on stellar mass}
\label{sec:MRmass}

\begin{figure}
\includegraphics[width=0.99\columnwidth, trim = 0 15 0 0]{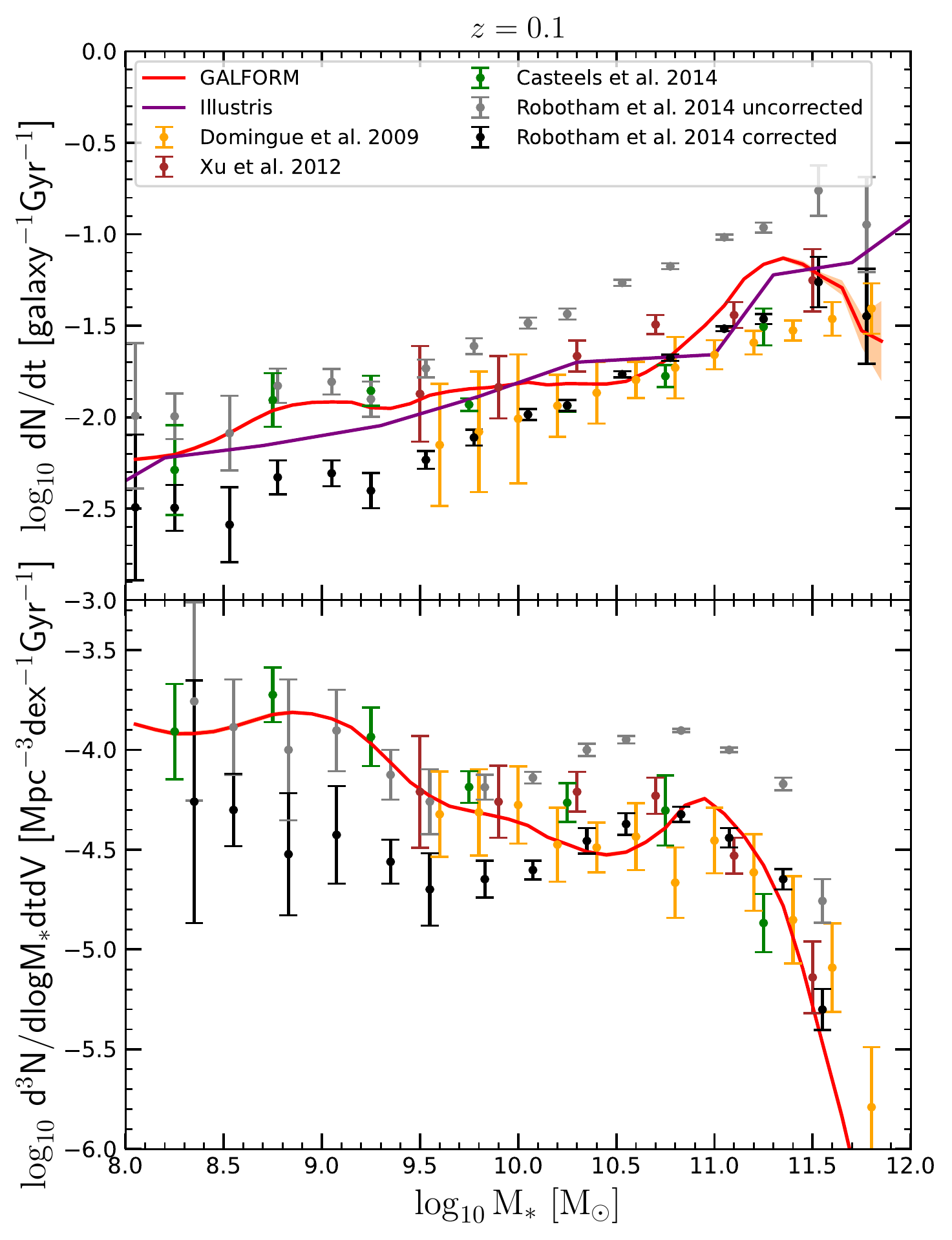}
\caption{Major ($\mu_* \in [0.25,1]$) galaxy merger rate as a function of stellar mass at redshift $z\approx0.1$. Error bars and shaded regions correspond to $1\sigma$-confidence intervals. The uncorrected data from Robotham et~al.\protect\cite{Robotham} refers to their standard sample, while the corrected data includes corrections for visual disturbances (see text for details). \textit{Top}: Merger rate per galaxy compared with observations and Illustris simulation (Rodriguez-Gomez et~al.\protect\citealt{RodrGomez}). \textit{Bottom}: Merger rate density compared with observations.}
\label{fig:figMRMass}
\end{figure}

The top panel of Fig.~\ref{fig:figMRMass} shows our predicted major ($\mu_* \in [0.25,1]$) merger rate per galaxy as a function of stellar mass at redshift $z\approx0.1$. The merger rate agrees well with observations, as well as with the Illustris simulation (Vogelsberger et~al. \citealt{Vogelsberger}), for which we take merger rates from Rodriguez-Gomez et~al.~\cite{RodrGomez}. Good agreement is found with observational data from Domingue et~al.~\cite{Domingue} (SDSS) and Xu et~al.~\cite{Xu} (COSMOS). Our major merger rate agrees well with that of Casteels et~al.~\cite{Casteels}, although we note that this comparison is somewhat moot since their results are based on measurements of merger fractions.

The uncorrected and corrected (for visual disturbances) measurements of merger rates found by Robotham et~al.~\cite{Robotham}, using GAMA-II, generally thread our prediction. We note that a comparison should in principle be done with the uncorrected version of their results since these do not use any morphological information. Interestingly, the correction for visual disturbances brings their measurements much more in line with other observations and our results, even though it does not represent a physically motivated correction to close pair counts (if one uses merger timescales which account for spurious pairs, e.~g. the KW08 one). We reach the same conclusion in terms of the dependence of their merger rate on redshift (\S~\ref{MRredshift}), as well as when directly comparing their close pair fractions with those from other studies (\S~\ref{sec:CPF}).

One feature of our predicted merger rate per galaxy, which has to our knowledge not been predicted by any other model and is only hinted at in the Robotham et~al.~\cite{Robotham} measurements, is the turnover at high masses. It is possible that other simulations and observational studies have smoothed out this feature due to the limited volumes probed. Our major and minor merger rates shown in Fig.~\ref{fig:figMRMass} are based on  $\approx16$ million merger events (spread over 40 mass bins between $M_*=10^{8}$ M$_\odot$ and $M_*=10^{12}$ M$_\odot$).

\begin{figure}
\includegraphics[width=0.99\columnwidth, trim = 0 15 0 0]{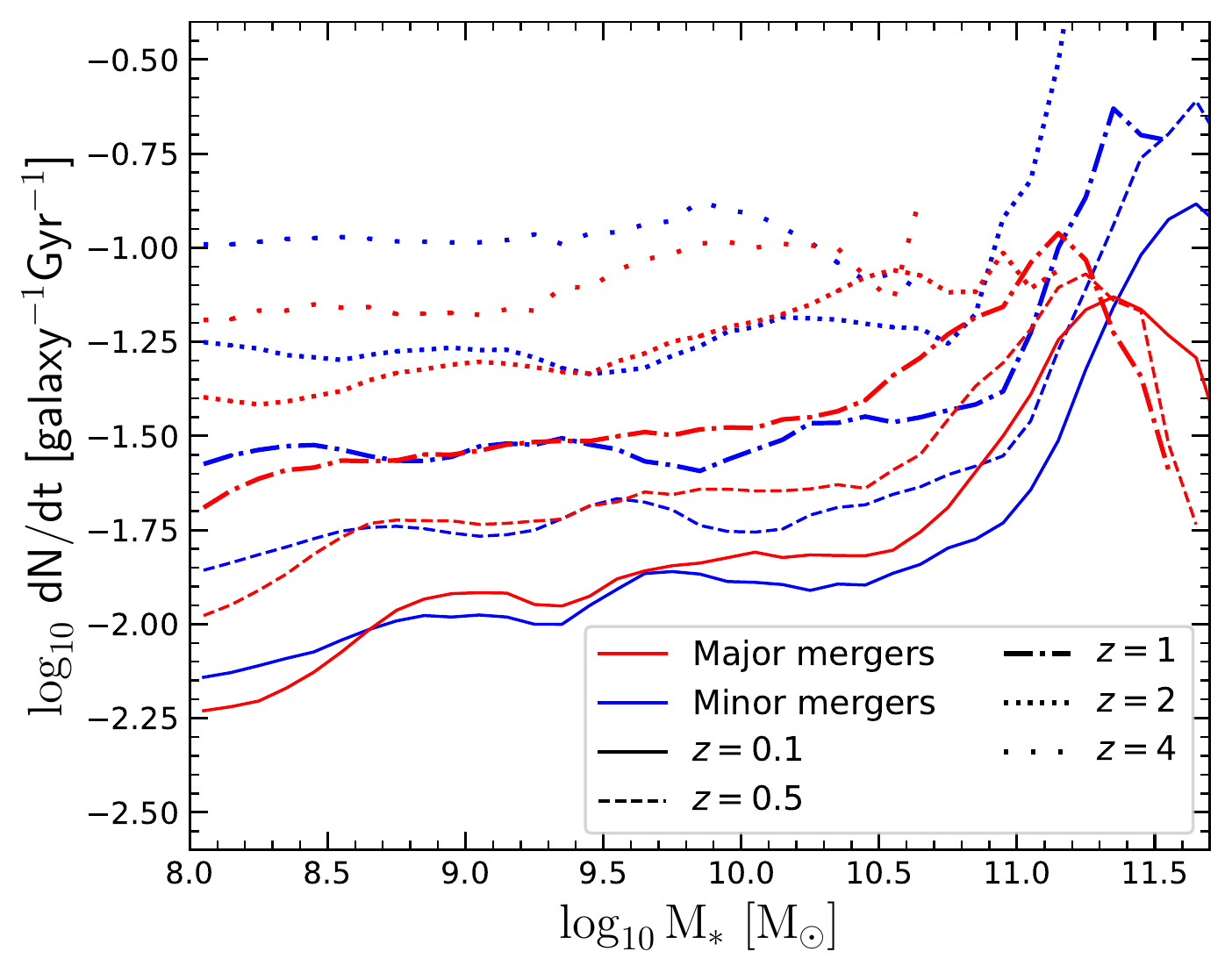}
\caption{Merger rates of galaxies for minor ($\mu_* \in [0.1,0.25]$; blue) and major ($\mu_* \in [0.25,1]$; red) mergers as functions of stellar mass (of the most massive progenitor) at various redshifts. Different line types represent different redshifts, as indicated by the legend. Lines are discontinued at points beyond which no mergers were found.}
\label{fig:figMRMDiffR}
\end{figure}

It might be argued that this turnover in the merger rate is a flawed prediction of {\tt GALFORM}. Perhaps counter-intuitively, the turnover suggests that the most massive galaxies (likely to be either members or the central galaxies of clusters) have slower merger growth than less massive counterparts, despite this being their main mode of growth. However, there are theoretical argument which boost our confidence that this is a genuine, physical feature which will be confirmed by larger observational studies. 
An important feature that determines the merger rate is the mass distribution of satellites below $M_*$. This corresponds to the stellar mass function (GSMF) of satellite galaxies, which inherits most of the features of the overall GSMF. Specifically, the satellite GSMF also displays an exponential drop at high masses (e.g. Yang et~al. \citealt{Yang}, Tal et~al. \citealt{Tal}, Weigel et~al. \citealt{Weigel}). For galaxies with $M_*=10^{11}$ M$_\odot$ (the start of the exponential drop in the GSMF at $z=0$, Fig.~\ref{fig:figSMF}), we expect that the major merger rate above this mass will begin to change behaviour due to an exponentially decreasing number of companions of comparable mass. This agrees with the start of the turnover in the major merger rate in Fig.~\ref{fig:figMRMass}. 

The predicted merger rate density, shown in the bottom panel of Fig.~\ref{fig:figMRMass}, is found to agree well with observational studies. There is a small discrepancy at lower masses, but this is well within the observational uncertainty. This is in part due to {\tt GALFORM} predicting a larger number of low mass galaxies than is observed.

\begin{figure*}
\includegraphics[width=\textwidth, trim = 0 15 0 0]{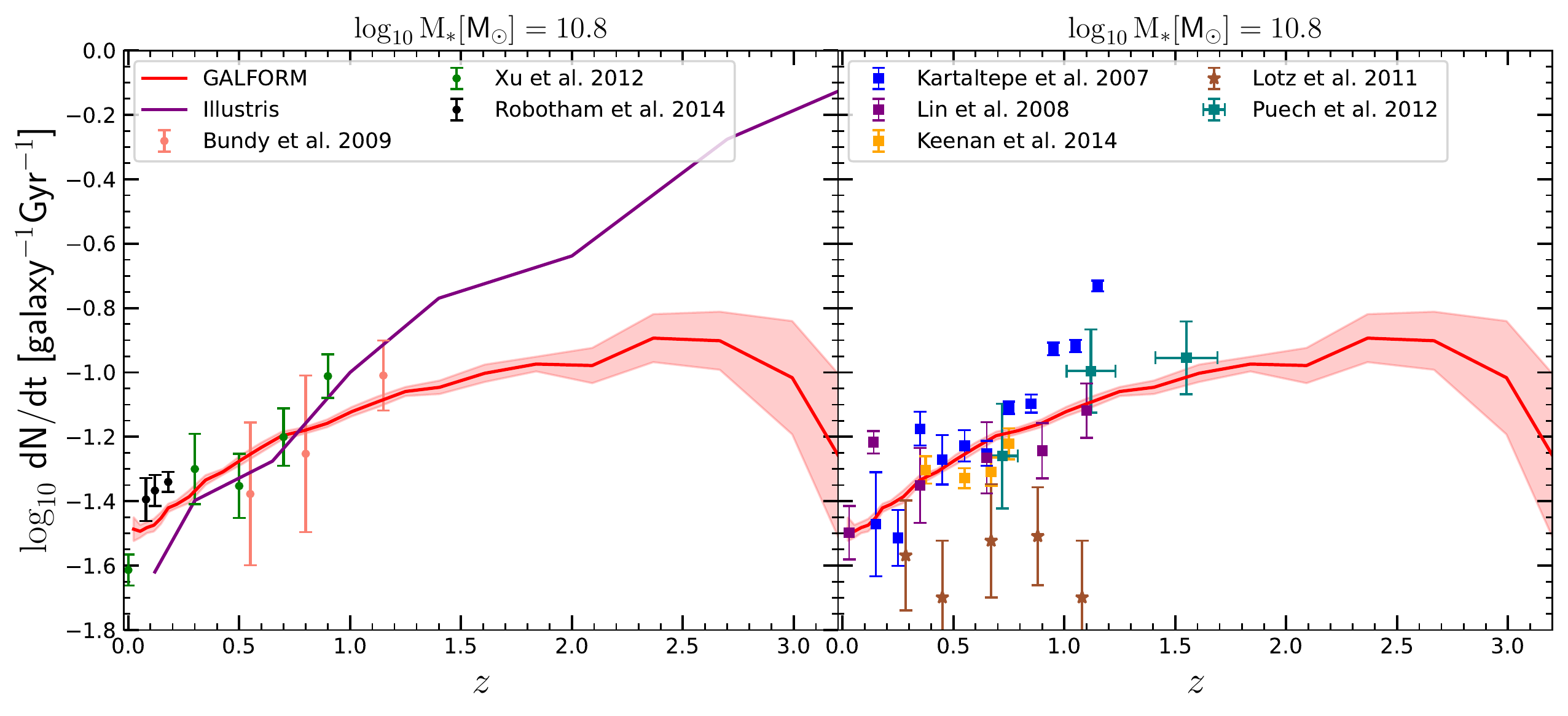}
\caption{Major ($\mu_* \in [0.25,1]$) merger rate per galaxy for $M_*=10^{10.8}$ M$_\odot$. Symbols with error bars show observational estimates as given by the legend. Error bars and shaded regions show $1\sigma$-confidence intervals. {\it Left:} Our predictions compared with observational merger rates obtained by use of close pair fractions, selected in line with what is described in \S~\ref{sec:ObsTimeScale}, and those of the Illustris simulation (purple line, Rodriguez-Gomez et~al.\protect\citealt{RodrGomez}). Merger rates were derived from observational studies by applying a merger timescale from KW08 to measured close pair fractions, which were also multiplied by factors to account for different selection criteria (see Table~\ref{tab:tab0}, \S~\ref{sec:ObsTimeScale} and \ref{sec:ObsDiscr} for details). {\it Right:} Our predictions compared with merger rates obtained through close pair fractions selected using a luminosity definition of major mergers ($L_{\mathrm{sec}}/L_{\mathrm{pri}}>0.25$, squares), or those obtained by use of merger fractions (stars). This shows the need for care when comparing merger rates obtained through different methods.}
\label{fig:figMRRObs108}
\end{figure*}

Parry et~al.~\cite{Parry} previously investigated the role of mergers in the buildup of galaxy spheroids, using two different semi-analytical models; {\tt GALFORM} and L-GALAXIES. They found that only the most massive spheroids ($M_*>10^{11.3}$ M$_\odot$) were built through major mergers. Most other spheroids were built primarily through minor mergers and disc instabilities, with most galaxies never experiencing a major merger. Our Fig. \ref{fig:figMRMDiffR} shows that \st{in the new version of {\tt GALFORM} (with an updated merging scheme)} minor mergers are as frequent as major mergers over a wide range of masses and redshifts. For $M_*>10^{11.3}$ M$_\odot$, at $z=0$, minor mergers overtake major ones in frequency. At higher redshifts, minor mergers overtake major ones at even lower masses ($M_*=10^{11}$ M$_\odot$ at $z=2$) . Whether this is consistent with minor mergers overtaking major ones in terms of mass growth, as seen in observations (e.g. Ownsworth et~al. \citealt{Ownsworth}), can only be confirmed by studying mass growth rates due to mergers. We plan to investigate this in a future paper.

Note that the relative frequency of major and minor mergers is very sensitive to the limiting mass ratio $\mu_*$. In this work we have used $\mu_*=1/4$. However, using $\mu_*=1/3$, as is done in many other works, results in an increase of a factor of $\approx1.65$ in the ratio between major and minor merger rates (based on the $f_\mathrm{mg}-\mu_*$ relation, see Section \ref{sec:CPmethod}).

\subsection{Dependence on redshift}
\label{MRredshift}

The left panel of Fig.~\ref{fig:figMRRObs108} shows major merger rates per galaxy at $M_*=10^{10.8}$ M$_\odot$ as a function of redshift up to $z=3.2$, compared with observational estimates (up to $z=1.2$) and the Illustris simulation (although we note that their results are for $M_*=10^{11}$ M$_\odot$). We see that both {\tt GALFORM} and Illustris agree roughly with observational estimates plotted here. {\tt GALFORM} agrees better with Robotham et~al.~\cite{Robotham}, while Illustris agrees better with Xu et~al.~\cite{Xu}. Measurements from Bundy et~al.~\cite{Bundy} are consistent with both models. Beyond $z\approx1$, the predictions from the two models diverge significantly. Illustris predicts a rising merger rate, while in {\tt GALFORM} a turnover is clearly seen. Merger rates from Illustris agree slightly better with the last observational data point from Xu et~al. \citealt{Xu}.

In the right panel of Fig.~\ref{fig:figMRRObs108}, we also show merger rates obtained through the use of pair selections which are not necessarily in line with our standard definition, described in \S~\ref{sec:ObsDiscr}. In particular, we show measurements obtained by use of close pair fractions for major mergers defined through a pair luminosity threshold of $L_{\mathrm{sec}}/L_{\mathrm{pri}}>0.25$, by Kartaltepe et~al. \cite{Kartaltepe}, Lin et~al. \cite{Lin} and Keenan et~al. \cite{Keenan}. Merger rates obtained in this way are similar to those that result from using a standard stellar mass threshold. This is not surprising, since the rest of the method (selecting pairs within some separation) is the same. However, Mantha et~al.~\cite{Mantha} show that pair fractions (and thus merger rates) measured in this way tend to show less of a plateau with redshift compared to ones selected through stellar mass. This can indeed be seen in the measurements by Karteltepe et~al.~\cite{Kartaltepe}, with no sign of a plateau. We also show merger fraction measurements from Lotz et~al.~\cite{Lotz2011}, which include a morphological selection. The deviation from our predictions and other measurements shows that this kind of comparison is even more uncertain. Finally, we show measurements based on both morphological and kinematical data, by Puech et al. \cite{Puech}, which seem consistent with GALFORM predictions.

\begin{figure*}
\includegraphics[width=0.95\textwidth, trim = 0 15 0 0]{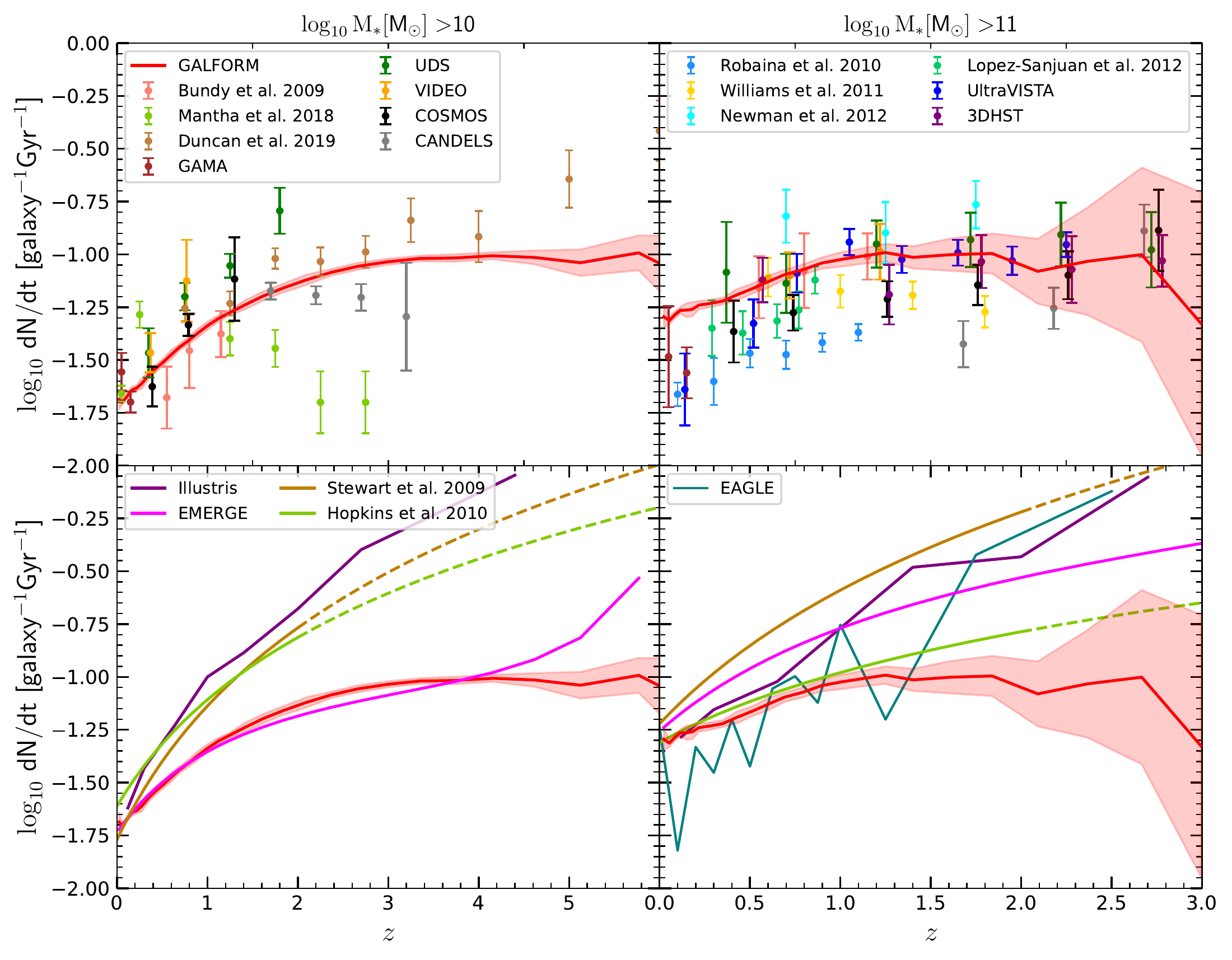}
\caption{Merger rate per galaxy for galaxies above stellar mass thresholds (shown above each panel) as functions of redshift. Red lines show our major ($\mu_* \in [0.25,1]$) merger rate, while dots give results from various observational studies, as given by the key. UltraVISTA and 3DHST results are from Man et~al.\protect\cite{Mundy}, while other named survey field results are from Mundy et~al.\protect\cite{Mundy}. Coloured lines give results from other models, as per the legend. Dashed lines indicate extrapolations of model merger rates. Observational merger rates were obtained from close pair fractions by division with a universal KW08 timescale, as well as correcting for  different selections (see Table~\ref{tab:tab0}, \S~\ref{sec:ObsTimeScale} and \ref{sec:ObsDiscr}). Error bars and shaded regions correspond to $1\sigma$-confidence intervals.}
\label{fig:figMRRObsLar}
\end{figure*}

The comparison from the left panel of Fig.~\ref{fig:figMRRObs108} shows that it is not clear if {\tt GALFORM} predicts a correct evolution of the merger rate with redshift. We now turn to a comparison using a threshold mass selection, for which there are much more extensive data available. This allows probing merger rates up to higher redshifts. We compare our results with merger rates obtained through our standard close pair definition (\S~\ref{sec:ObsDiscr}).  These comparisons are made for two stellar mass thresholds, $M_*>10^{10}$ M$_\odot$ and $M_*>10^{11}$ M$_\odot$. In the top panels of Fig.~\ref{fig:figMRRObsLar} we show our predictions in comparison with various observational studies, as indicated in the legend. The details of these studies can be found in Table~\ref{tab:tab0}.

The top left panel of Fig.~\ref{fig:figMRRObsLar} shows the comparison for the selection $M_*>10^{10}$ M$_\odot$. {\tt GALFORM} shows rough agreement with many of the observational data points, but also disagrees significantly with Mantha et~al.~\cite{Mantha} at low redshifts ($z>2$), and slightly with Duncan et~al.~\cite{Duncan} and many measurements by Mundy et~al.~\cite{Mundy}. However, these observational measurements also significantly disagree among each other.  This is especially disconcerting given that some studies calculated close pair fractions from the same field (e.g. Mantha et~al. \citealt{Mantha} and Duncan et~al. \citealt{Duncan} both studied galaxy pairs in CANDELS). Furthermore, measurements by Mundy et~al.~\cite{Mundy} in different survey fields also show different trends (e.g UDS and VIDEO points implying a monotonic rise with redshift, and CANDELS points showing a plateau).

In the top right panel of Fig.~\ref{fig:figMRRObsLar} we show the corresponding comparison for $M_*>10^{11}$ M$_\odot$.  {\tt GALFORM} agrees particularly well with the measurements from Bundy et~al.~\cite{Bundy}, Man et~al.~\cite{Man} and Mundy et~al.~\cite{Mundy}. However, there is disagreement with a slew of other measurements, which show a consistently lower normalisation than the ones mentioned so far. Specifically, the GAMA and CANDELS measurements of Mundy et~al.~\cite{Mundy}, the low-redshift measurement from UltraVISTA by Man et~al.~\cite{Man}, and the studies by Robaina et~al.~\cite{Robaina}, Williams et~al.~\cite{Williams} and Lopez-Sanjuan et~al.~\cite{LopezSanjuan2012} are consistent with merger rates up to a factor of two lower than most other observations and in {\tt GALFORM}. Newman et~al.~\cite{Newman}, on the other hand, measure merger rates that are somewhat larger than the trends from any other studies.

We also compare our results with those from three semi-empirical models (Stewart et~al. \citealt{Stewart}, Hopkins et~al. \citealt{Hopkins} and {\tt EMERGE}: O'Leary et~al. \citealt{OLeary}), as well as the Illustris (Rodriguez-Gomez et~al. \citealt{RodrGomez}) and EAGLE (Lagos et~al. \citealt{Lagos}) hydrodynamical simulations. All three of the mentioned semi-empirical models are based on populating N-body simulation dark matter haloes with galaxies. Stewart et~al.~\cite{Stewart} and Hopkins et~al.~\cite{Hopkins} do this by means of (sub)halo abundance matching, while in {\tt EMERGE}, an instantaneous star formation efficiency is used to obtain the correct galaxy abundances and stellar masses. In neither the Stewart et~al.~\cite{Stewart} nor Hopkins et~al.~\cite{Hopkins} model are subhaloes tracked while they evolve inside a primary halo; instead, a merger time is set as soon as a halo becomes a subhalo of a larger halo (as in older versions of {\tt GALFORM}). In {\tt EMERGE}, subhaloes are treated in the same way as in {\tt GALFORM}.
All three models use dynamical friction merger times derived by Boylan-Kolchin et~al.~\cite{Boylan2009}, which are different to ours (Eqn.~\ref{eq:TmergSimha}). These differences can potentially lead to large disagreements. Furthermore, the results from Stewart et~al.\cite{Stewart} and Hopkins et~al.\cite{Hopkins} are based on abundance matching up to $z=2$, so merger rates predicted by these models beyond that redshift should be treated as extrapolations.

We show the comparison between {\tt GALFORM} and different theoretical models in the bottom panels of Fig.~\ref{fig:figMRRObsLar}. For the $M_*>10^{10}$ M$_\odot$ mass selection, {\tt GALFORM} agrees best with {\tt EMERGE}. However, {\tt GALFORM} predicts a plateau in the merger rate, whereas {\tt EMERGE} predicts an ever-rising merger rate beyond $z=4$ (although this is possibly a result of their sample size restriction at high redshifts). The results from Illustris, Stewart et~al.\cite{Stewart} and Hopkins et~al.\cite{Hopkins} all predict merger rates which rises much faster than either {\tt GALFORM} or {\tt EMERGE}.

For the $M_*>10^{11}$ M$_\odot$ mass selection (bottom right panel of Fig.~\ref{fig:figMRRObsLar}), all six theoretical models show remarkable agreement for $z=0$, but they disagree at higher redshifts. All models show a rise in the merger rate at all redshifts shown, unlike {\tt GALFORM}, which features a plateau and turnover (although the turnover is in the regime in which our merger rate is uncertain due to sample size restrictions). Illustris, EAGLE and Stewart et~al.\cite{Stewart} agree fairly well in predicting a steep rise. {\tt EMERGE} and Hopkins et al. \cite{Hopkins} predict a shallower rise, with the results of Hopkins et al.\cite{Hopkins} in better agreement with {\tt GALFORM}.

It should be noted that the predictions from Illustris, if taken in a broader context, are somewhat puzzling. Matching the close pair fractions from Illustris to the corresponding merger rates, one can infer a merger timescale. This procedure results in a merger timescale that evolves as $T_\mathrm{mg}\propto(1+z)^{-2}$ (Snyder et~al. \citealt{Snyder}), which is in clear disagreement with most other results (KW08, Lotz et~al. \citealt{Lotz2010b}, Jiang et~al. \citealt{Jiang2014}, O'Leary et~al. \citealt{OLeary}). We note that merger rates in hydrodynamical simulations depend sensitively on the way they are defined. Specifically, the moment at which the mass ratio $\mu$ of two galaxies is calculated can affect the merger rate significantly, as shown in Rodriguez-Gomez et~al.~\cite{RodrGomez}. It is possible that a more suitable method of calculating the merger mass ratio might reduce this disagreement. Whether the merger rate is defined in terms of progenitor or descendant mass can also have a significant impact (Rodriguez et al. \citealt{RodrGomez}, O'Leary et al. \citealt{OLeary}).

\begin{figure}
\includegraphics[width=\columnwidth,trim = 0 15 0 0]{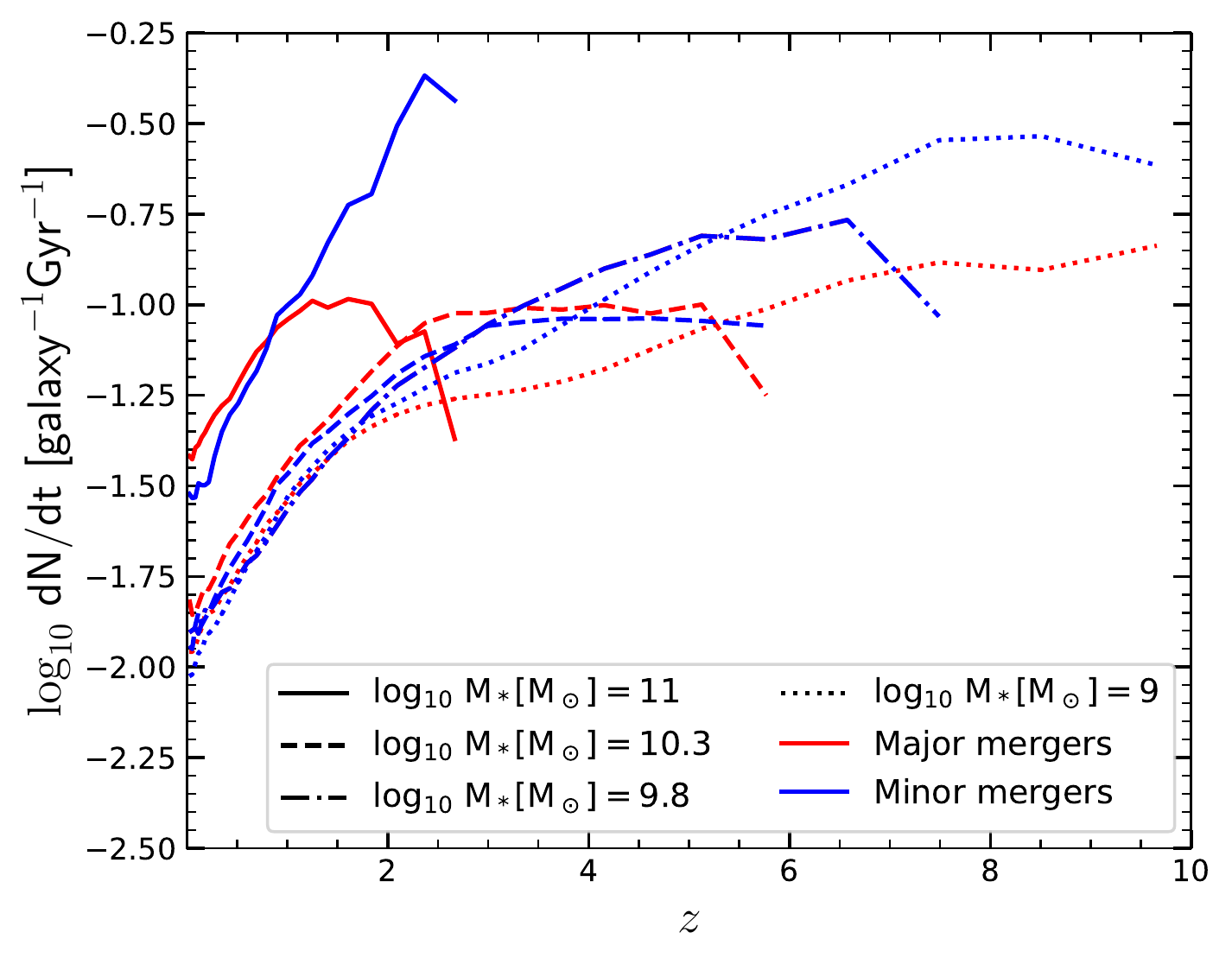}
\caption{Minor ($\mu_* \in [0.1,0.25]$) and major ($\mu_* \in [0.25,1]$) merger rates per galaxy for galaxies of several masses (as given by the legend) as functions of redshift. Lines are discontinued at points beyond which no mergers were found.}
\label{fig:figMRRDiffM}
\end{figure}

Fig.~\ref{fig:figMRRDiffM} shows the evolution of the major and minor merger rates per galaxy, as predicted by {\tt GALFORM}, for several different masses. The merger rates increase up to some redshift after which they drop off rapidly. This turnover redshift decreases with increasing mass. The reduction in merger rates with redshift is related to that seen in the merger rate per galaxy as a function of stellar mass (top panel of Fig.~\ref{fig:figMRMDiffR}). The merger rates drop off beyond some redshift because this is the redshift where that mass passes into the exponentially decreasing regime in the stellar mass function (Fig.~\ref{fig:figSMF}).




\section{Close pair fraction}
\label{sec:CPF}

While merger rates are a useful quantity for galaxy formation models, they cannot be measured directly. This is because a merger timescale must be assumed, and they are always obtained from theoretical models. Here we calculate the close pair fraction of galaxies, as described in \S~\ref{sec:CPmethod}. We study the major ($\mu_* \in [0.25,1]$) close pair fraction as a function of stellar mass and redshift. We leave an investigation of its dependence on the maximum projected separation $r_\mathrm{max}$ and velocity separation $v_\mathrm{max}$ for \S~\ref{sec:TSres}, where we analyse the merger timescale (note that the merger timescale and pair fraction depend on $r_\mathrm{max}$ and $v_\mathrm{max}$ in the same way). 

For simplicity, and in order to match recent observational work, we focus on close pairs within a projected separation of $r_\mathrm{max}=20$ $h^{-1}$kpc and a relative line-of-sight velocity less than $v_\mathrm{max}=500$ kms$^{-1}$. When comparing our model predictions with results from studies using other selections, we apply conversions as described in Section \ref{sec:ObsDiscr}, and given in \S~\ref{sec:TSres}.

\begin{figure*}
\includegraphics[width=\textwidth,trim = 0 15 0 0]{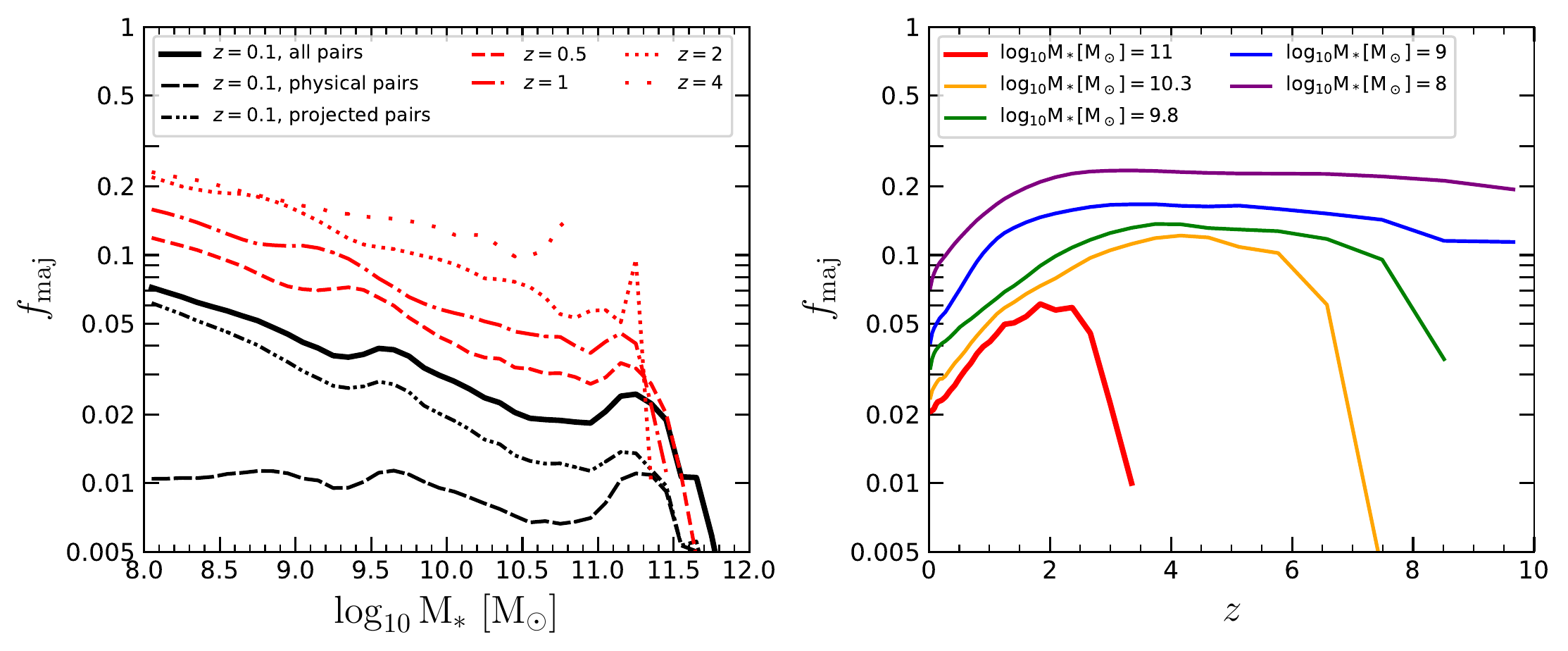}
\caption{Close pair fraction of major ($\mu_* \in [0.25,1]$) galaxy pairs with pair selection criteria of $r_\mathrm{sep}<20$ $h^{-1}$kpc and $\vert v_\mathrm{sep}\vert<500$ kms$^{-1}$. Lines are discontinued at masses or redshifts beyond which no close pairs were found. {\it Left}: Major close pair fraction as a function of stellar mass at several redshifts. Black lines represent pair fractions for physical pairs (3D separations less than $20$ $h^{-1}$kpc), projected pairs (2D separations less than $20$ $h^{-1}$kpc, but 3D separations larger than $20$ $h^{-1}$kpc) or all pairs, as per the legend. ${\it Right}$: Major close pair fraction as a function of redshift for several stellar masses.}
\label{fig:figCPMR}
\end{figure*}

The left panel of Fig.~\ref{fig:figCPMR} shows the major close pair fraction as a function of stellar mass at different redshifts. The close pair fraction is generally a decreasing function of stellar mass. However, its behaviour is not entirely straightforward to understand, as it includes both physical pairs (with 3D separations comparable to their projected separation) and projected pairs (with line-of-sight separations much larger than their projected separations). We have split up the two contributions for $z=0.1$ in Fig.~\ref{fig:figCPMR}. This shows that physical pair fractions (akin to merger fractions) are almost constant with mass, whereas projected pair fractions decrease with mass. The latter is expected from the decreasing behaviour of the stellar mass function. Even at large stellar masses, however, around half of all pairs come from projection. In these massive systems, these projected pairs are almost exclusively located in the same dark matter halo. We have kept projected close pairs in our analysis since they cannot be separated from physical pairs in observational studies.

For  $M_*>10^{11}$ M$_\odot$ the close pair fraction reaches a maximum and turns over at higher masses. This happens for the same reason as the merger rate (Section \ref{sec:MRmass}), and this drop is also seen in observational results from Robotham et~al.~\cite{Robotham}. The turnover at high masses shifts to lower masses with increasing redshift, similar to what is seen for the merger rate (Fig.~\ref{fig:figMRMDiffR}). Again, this is due to galaxies entering the exponentially decreasing regime of the stellar mass function.

The right panel of Fig.~\ref{fig:figCPMR}  shows how major close pair fractions change with redshift at several masses. At high stellar masses ($M_*>10^{10}$ M$_\odot$) the close pair fraction shows a strong turnover, although this is sometimes hard to identify  due to a lack of galaxies at these redshifts. The turnover is also present for less massive galaxies, but is much weaker. The presence of a turnover for all stellar masses is harder to confirm for merger rates (Fig.~\ref{fig:figMRRDiffM}) since we generally detect a smaller number of mergers than we do close pairs for any mass bin and redshift. Our explanation for this turnover remains the same: it is the result of the behaviour of the stellar mass function.

\subsection{Comparison with observations and other models}
\subsubsection{Mass dependence}

\begin{figure}
\includegraphics[width=0.99\columnwidth, trim = 0 15 0 0]{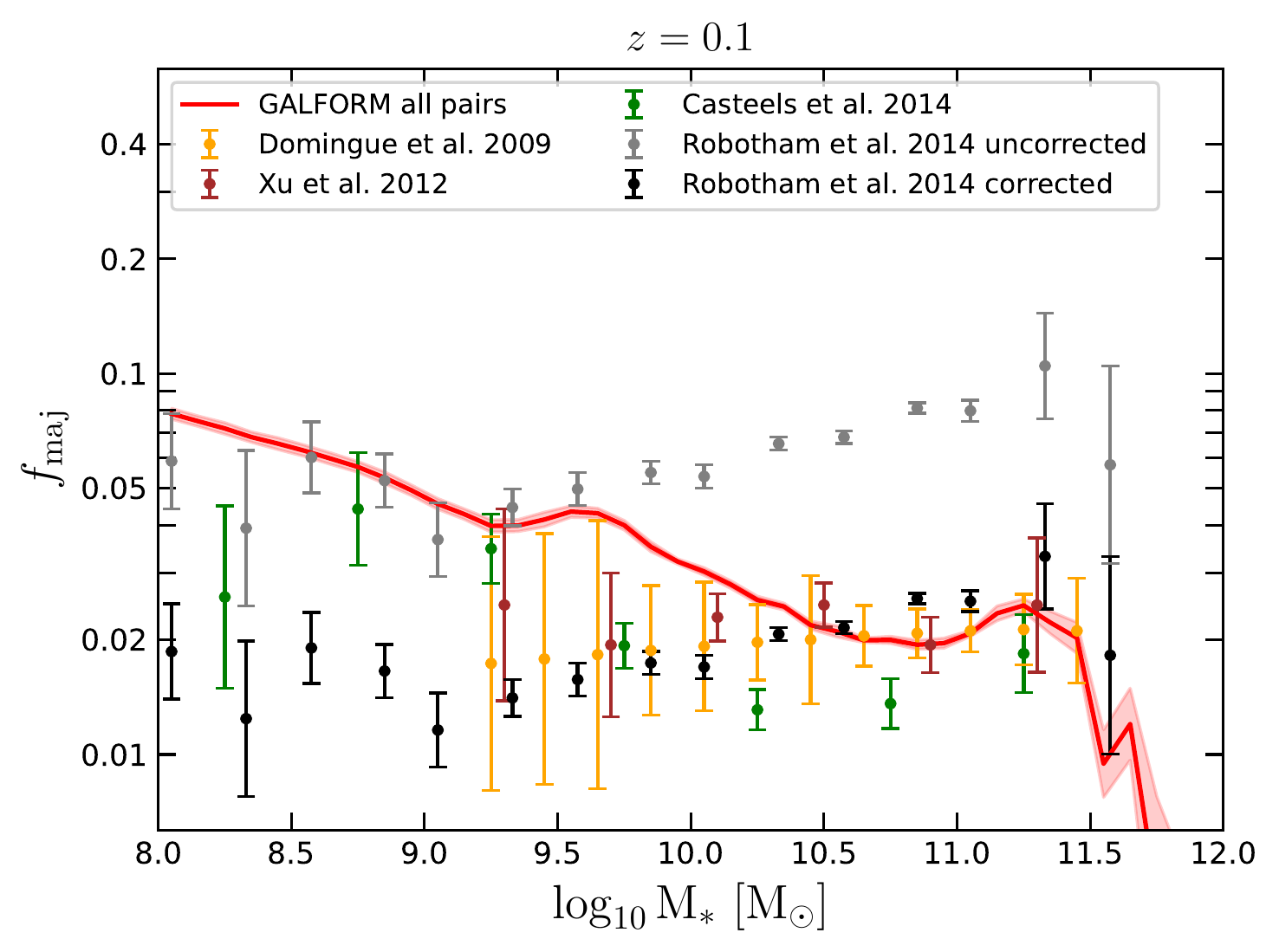}
\caption{Close pair fraction of major ($\mu_* \in [0.25,1]$) galaxy pairs from our analysis (red line and shaded region) compared with observations as a function of stellar mass, at $z=0.1$. The pair selection criteria applied are $r_\mathrm{sep}<20$ $h^{-1}$kpc and $\vert v_\mathrm{sep}\vert<500$ kms$^{-1}$. Results from observational studies were scaled up or down if their pair selection criteria are different from ours (see Table~\ref{tab:tab0} and \protect\S~\ref{sec:ObsDiscr} for details). Uncorrected results from Robotham et~al.~\protect\cite{Robotham} represent their standard sample, while their corrected data is that where corrections for visual disturbances were applied. Error bars and shaded regions represent $1\sigma$-confidence intervals.}
\label{fig:figCPObsM}
\end{figure}

Fig.~\ref{fig:figCPObsM} shows the fraction of major close pairs as a function of stellar mass, compared with the same observational datasets as considered for merger rates (Domingue et~al. \citealt{Domingue}, Xu et~al. \citealt{Xu}, Robotham et~al. \citealt{Robotham}, Casteels et~al. \citealt{Casteels}). Only observational studies which do not apply any additional selection (such as asymmetry cuts) are included with the exception of Casteels et~al.~\cite{Casteels}. We have included this study in the comparison since it is one of the more recent ones where the close pair fraction was studied as a function of stellar mass. 

Our close pair fraction is in fairly good agreement with observations for $M_*>10^{10}$ M$_\odot$. At lower masses the prediction diverges from the corrected results of Robotham et~al.~\cite{Robotham}, and those of Domingue et~al.~\cite{Domingue} and Xu et~al.~\cite{Xu}. We note that the latter two studies do not give results for $M_*<10^{9.5}$ M$_\odot$, and our predictions are within their range of uncertainty at the edge of this mass regime. At the same time, our results at low masses agree with the uncorrected results of Robotham et~al.~\cite{Robotham}, as well as those from Casteels et~al.~\cite{Casteels}. It should be noted that {\tt GALFORM} predicts a somewhat too large number of low-mass galaxies (see discussion in \S~\ref{sec:Intro}). We would expect this to be reflected as too large a close pair fraction, especially in the regime where projected pairs dominate the close pair fraction.


\subsubsection{Redshift dependence}

Most observational studies of close pair fractions examine its redshift dependence for a given mass (or above some threshold mass). We compare our results with the studies used earlier to compare the merger rate (see \S~\ref{sec:MergRates}), and also with observational studies of close pairs from the MUSE fields (Ventou et ~al. \citealt{Ventou2019}, with improved methodology and expanded datasets compared to Ventou et~al.~\citealt{Ventou2017}), as well as other theoretical studies. Close pair fractions have been studied in the Illustris simulation (Vogelsberger et ~al. \citealt{Vogelsberger}) and the {\tt EMERGE} (Moster et~al. \citealt{Moster}) and {\tt UNIVERSEMACHINE} (Behroozi et~al. \citealt{Behroozi}) semi-empirical models. The pair fractions from these models were taken from Snyder et~al.~\cite{Snyder}, O'Leary et~al.~\cite{OLeary} and Endsley et~al.~\cite{Endsley}, respectively. We do not compare our results with merger fractions from EAGLE (Qu et~al. \citealt{Qu}) or HorizonAGN (Kaviraj et~al. \citealt{Kaviraj}), since these are inherently not comparable with close pair fractions.

Close pairs are usually taken from parent samples of galaxies whose stellar mass is chosen to be \textit{above} a given threshold value.
We choose the following close pair samples from our results for the purpose of comparison: $M_*>10^{9.5}$ M$_\odot$, $M_*>10^{10}$ M$_\odot$, $M_*=10^{10.8}$ M$_\odot$ and $M_*>10^{11}$ M$_\odot$. We compare results from studies with $M_*>10^{10.3}$ M$_\odot$ samples with our $M_*>10^{10}$ M$_\odot$ sample, while studies with $10^{10.5}<M_*<10^{11}$ M$_\odot$ are compared with our $M_*=10^{10.8}$ M$_\odot$ results. We do not expect these differences to be problematic since the close pair fraction does not vary strongly with mass in this mass regime (Fig.~\ref{fig:figCPMR}).

\begin{figure*}
\includegraphics[width=\textwidth, trim = 0 20 0 0]{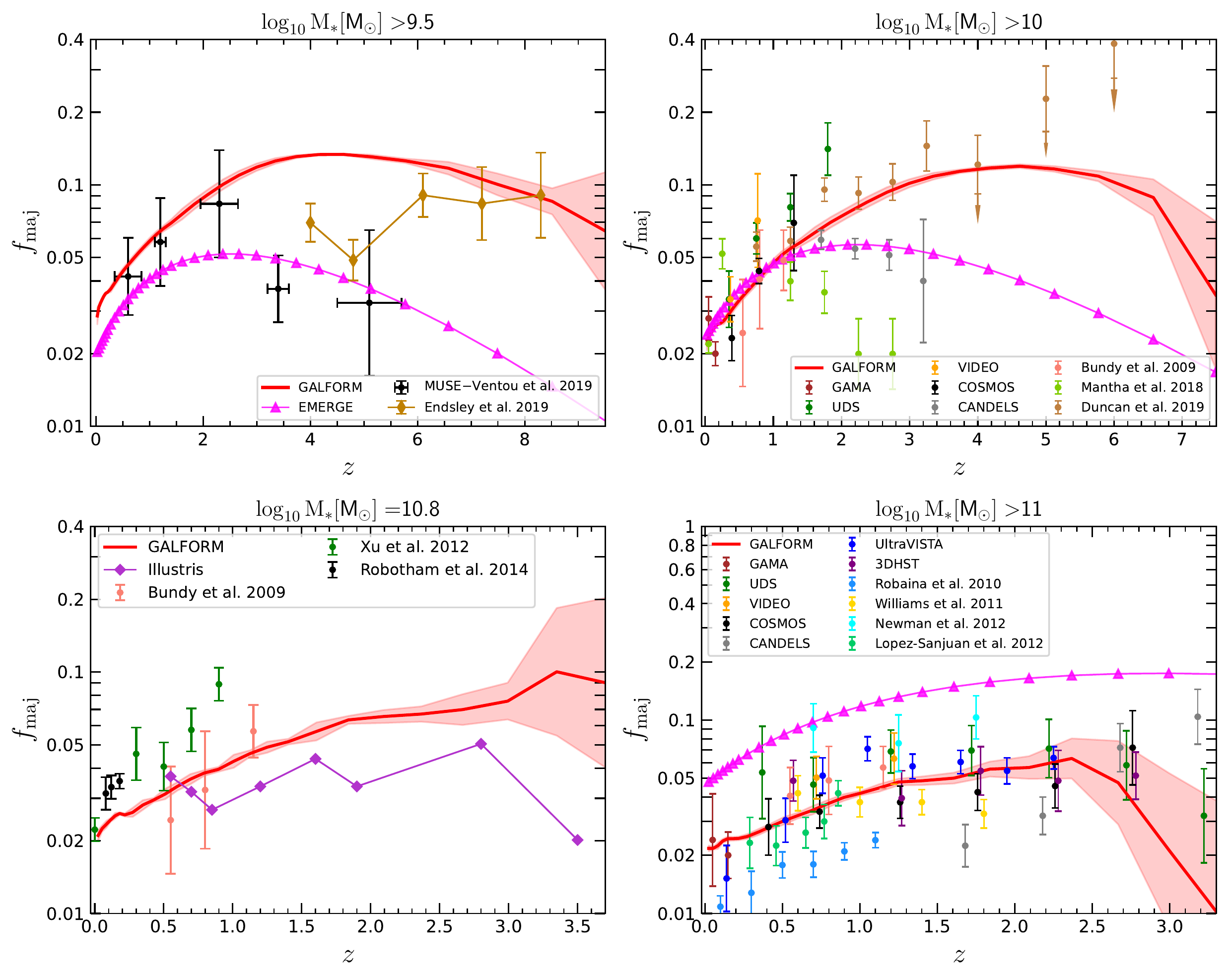}
\caption{Close pair fraction of major ($\mu_* \in [0.25,1]$) galaxy pairs from our analysis (red lines and shaded regions) compared with observations (symbols with error bars) and other theoretical models (symbols connected by lines) as a function of redshift. The pair selection criteria are $r_\mathrm{sep}<20$ $h^{-1}$kpc and $\vert v_\mathrm{sep}\vert<500$ kms$^{-1}$. Each plot represents a different mass selection, as shown above the panels. Results from observational and theoretical studies were scaled up or down if their selection criteria were different from ours (see Table~\ref{tab:tab0} and \S~\ref{sec:ObsDiscr}). UltraVISTA and 3DHST results are from Man et~al.\protect\cite{Man}, while other named survey field results are from Mundy et~al.\protect\cite{Mundy}. Error bars and shaded regions represent $1\sigma$-confidence intervals.}
\label{fig:figCPObsR}
\end{figure*}

The top left panel of Fig.~\ref{fig:figCPObsR} shows the close pair fraction for galaxies with $M_*>10^{9.5}$ M$_\odot$. At higher redshifts ($z>3$) the observational measurements by Ventou et~al.~\cite{Ventou2019} imply a sharper drop and a lower plateau than our prediction.  

The top right panel of Fig.~\ref{fig:figCPObsR} shows results for the mass selection $M_*>10^{10}$ M$_\odot$. Observational results from various studies show close agreement with {\tt GALFORM} for $z<2$. The exception is the second data point of the CANDELS results from Mantha et~al.~\cite{Mantha}, which show a much sharper rise to $f_\mathrm{maj}=0.1$ at $z\approx0.3$ and a similarly sharp decrease to $f_\mathrm{maj}=0.02$ by $z=2$. This drop is somewhat in agreement with the CANDELS results of Mundy et~al.~\cite{Mundy}, but even sharper, and in clear disagreement with the rising close pair fraction from Duncan et~al.~\cite{Duncan}. Our predictions agree with Duncan et~al.~\cite{Duncan} up to $z=4$, but are lower than their last two data points at $z=5$ and $z=6$, respectively. It should be noted that their values at these redshifts were inferred using incomplete information. Namely, the results from some of their fields are not well constrained, with only the upper bound of the close pair fraction determined. We have shown this effect as arrows pointing downwards for the last 3 data points, with the arrow size being in proportion to the number of fields exhibiting such results. 

The bottom left panel of Fig.~\ref{fig:figCPObsR} shows the comparison for $M_*=10^{10.8}$ M$_\odot$. For this mass selection, observational estimates are limited to $z<1.2$. Our predictions are lower than the observational data of  Robotham et~al.~\cite{Robotham} and Xu et~al.~\cite{Xu} over this redshift range, but similar to those of Bundy et~al.~\cite{Bundy}. 
It should be noted, however, that this mass selection corresponds to the regime in which our close pair fraction is expected to be somewhat too small since the same is true for the stellar mass function (Fig.~\ref{fig:figSMF}). 

In the bottom right panel of Fig.~\ref{fig:figCPObsR} we show predictions for $M_*>10^{11}$ M$_\odot$. In this mass regime, there is an impressive agreement between {\tt GALFORM} and most observational studies (Bundy et~al. \citealt{Bundy}, Lopez-Sanjuan et~al. \citealt{LopezSanjuan2012}, Man et~al. \citealt{Man}, Mundy et~al. \citealt{Mundy}). However, the measurements from Newman et~al.~\cite{Newman} are somewhat high, while those from Robaina et~al.~\cite{Robaina} and Williams et~al.~\cite{Williams} are somewhat low compared to the general trend. 

Finally, we compare our close pair fractions with those predicted by other theoretical models. Results from {\tt EMERGE} (O'Leary et~al. \citealt{OLeary}) are generally lower than ours for the first mass selection. For the second mass selection, we find agreement at low redshifts, but the {\tt EMERGE} pair fraction turns over quicker. For the highest mass selection ($M_*>10^{11}$ M$_\odot$), {\tt EMERGE} predicts a close pair fraction which is somewhat too large at all redshifts, compared with both observations and {\tt GALFORM}. This is somewhat surprising given the good agreement of merger rates seen in Fig. \ref{fig:figMRRObsLar}. These differences can be reconciled with different merger timescales (see \S \ref{sec:TSres} and Fig. \ref{fig:figTSComparison}. Endsley et~al.~\cite{Endsley} use the {\tt UNIVERSEMACHINE} semi-empirical model to create mock observations, in line with the capabilities expected of the \textit{JWST}. These results for the $M_*>10^{9.5}$ M$_\odot$ selection match those from {\tt EMERGE} and observations, and they also agree with the {\tt GALFORM} predictions at very high redshifts. Snyder et~al. \citealt{Snyder} have studied the close pair fraction in Illustris for galaxies with $10^{10.5}<M_*<10^{11}$ M$_\odot$. Their results are in fairly good agreement with ours at low redshifts, but are lower than both our predictions and observations for $z>1$.

Overall, by studying close pair fractions for different masses we have found that a coherent picture begins to emerge: close pair fractions flatten and start declining with increasing redshift for most mass selections. However, observations can differ dramatically, with some even predicting a  close pair fraction that increases monotonically with redshift. Theoretical models in general reproduce the behaviour found from most observations, but they also differ in detail. 

We note that our results on close pair fractions are much better constrained than any other theoretical predictions shown here. This is due to the fact that we have used the entire volume of the Planck Millennium simulation in our calculations. As a result, we have been able to provide predictions up to very high redshifts. The maximal redshifts are equal to 10, 7.5, 4.5 and 3.5 for the four mass selections that we consider.




\section{Merger timescale for close pairs}
\label{sec:TSres}

With both merger rates and close pair fractions calculated as functions of stellar mass and redshift, we are in position to derive the average merger timescale for samples of close pairs defined in the same way as in observational studies. This can be useful for inferring merger rates from observational measurements. Furthermore, by considering the dependence of the merger timescale on selection criteria, we can obtain formulas which can be used to convert close pair fractions from one selection to another. 

By definition (Eqn.~\ref{eq:eqCPTS}), the merger timescale for conversion of a close pair fraction $f$ to a merger rate per galaxy $\mathrm{d}N/\mathrm{d}t$ is 
\begin{equation}
T_\mathrm{mg}=f\times\bigg(\frac{\mathrm{d}N}{\mathrm{d}t}\bigg)^{-1},
\end{equation}
i.e. we only need to divide the close pair fraction by the merger rate to obtain the merger timescale. We first study how the merger timescale depends on stellar mass and redshift, and make a comparison with predictions from other models. Note that, unless specified, all results in this section are for samples of a given mass $M_*$, and not for samples with masses above a threshold value $M_*$. 

We apply a combination of 225 different close pair selection criteria to study how the merger timescale depends on the variables $r_\mathrm{max}$ and $v_\mathrm{max}$. Note that the dependencies of the merger timescale on these selection criteria are inherited from those for the close pair fraction. In particular, increasing $r_\mathrm{max}$ and $v_\mathrm{max}$ generally leads to more pairs, and thus a larger merger timescale (so that the merger rate remains the same, regardless of selection). For this reason, in this section we provide results only on how the merger timescale depends on $r_\mathrm{max}$ and $v_\mathrm{max}$, but the conclusions are exactly the same for the close pair fraction.

In order to motivate the fitting formulas we provide in this Section, as well as to make the results more transparent, we first study the dependence of the merger timescale on stellar mass and redshift for our standard selection ($r_\mathrm{max}=20$ $h^{-1}$kpc and $v_\mathrm{max}=500$ kms$^{-1}$). We then show how the merger timescale varies with $r_\mathrm{max}$, while $v_\mathrm{max}$ is kept fixed, and vice-versa. However, all parameters we give in this Section are obtained through a Markov Chain Monte Carlo fitting procedure, implemented through the {\tt emcee} Python package, and performed on the 4D grid of merger timescale values from the simulation (the variables being stellar mass, redshift, maximal separation and maximal velocity).

\subsection{Dependence on stellar mass and redshift}

\begin{figure*}
\includegraphics[width=\textwidth,trim = 0 15 0 0]{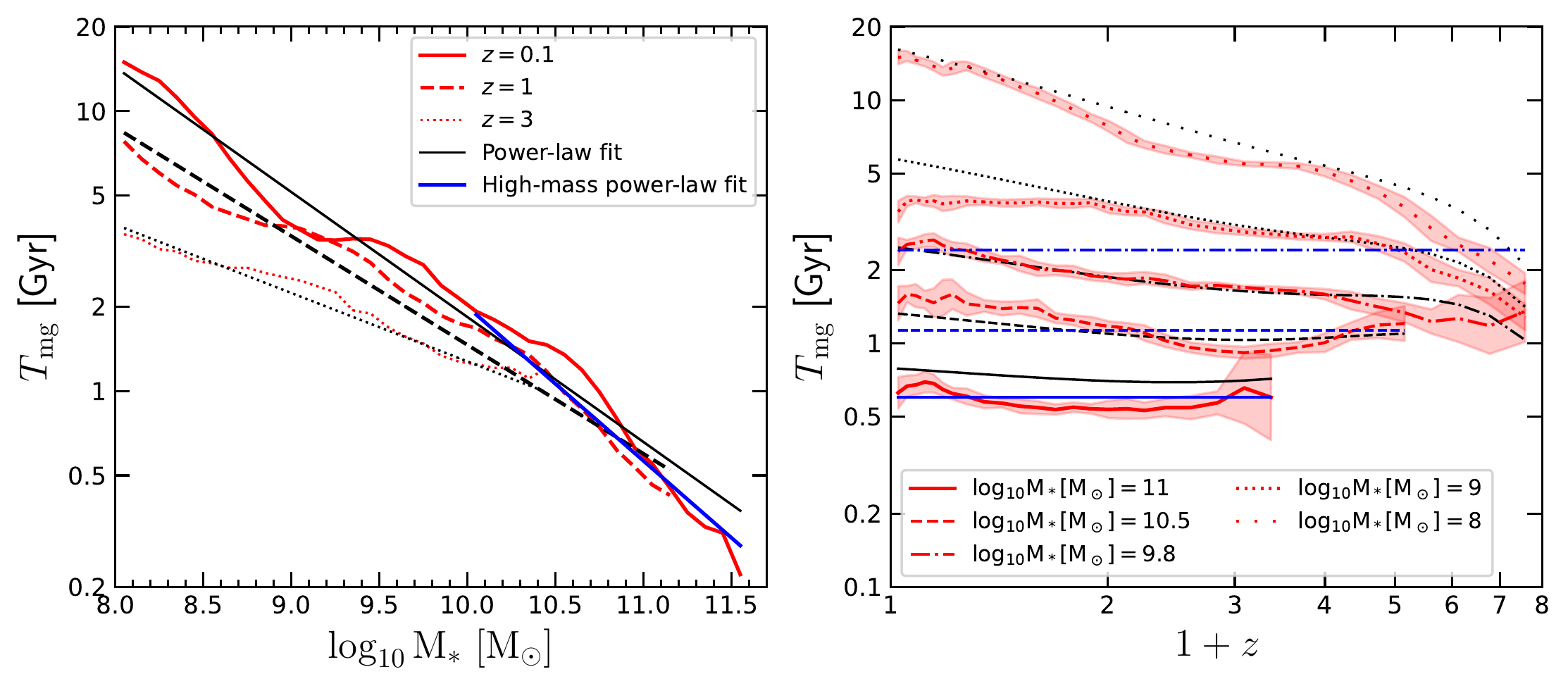}
\caption{Merger timescale for conversion of major ($\mu_* \in [0.25,1]$) close pair fractions to merger rates, with pair selection criteria of $r_\mathrm{sep}<20$ $h^{-1}$kpc and $\vert v_\mathrm{sep}\vert<500$ kms$^{-1}$. Lines are discontinued at masses or redshifts where the merger rate drops to zero (giving an infinite merger timescale) due to no detected merging galaxies. Red lines show the results from {\tt GALFORM}, while other lines correspond to fitting functions. Blue lines represent a simple, redshift-independent power law (Eqn.~\ref{eq:TSfitsimple}), while black lines represent a more complicated, but overall more accurate redshift-dependant power law (Eqn.~\ref{eq:TSfitfin}). {\it Left}: Merger timescale as a function of stellar mass. Black lines represent power law fits in stellar mass given by Eqn.~(\ref{eq:TSfitM}). {\it Right}: Merger timescale as a function of redshift for different masses. Black lines represent a fitting function based on power law fits in stellar mass, given by Eqn.~(\ref{eq:TSfitfin}).}
\label{fig:figTS}
\end{figure*}

Fig.~\ref{fig:figTS} shows the merger timescale predicted for major close pairs ($\mu_* \in [0.25,1]$) with pair selection criteria of $r_\mathrm{sep}<20$ $h^{-1}$kpc and $\vert v_\mathrm{sep}\vert<500$ kms$^{-1}$. In the left panel we show the merger timescale as a function of stellar mass. The timescale is approximately a power law in stellar mass. It decreases with stellar mass largely due to the diminishing number of projected pairs towards higher masses (which is itself a result of the stellar mass function), and due to the increase in the merger rate with mass. From the right hand panel we can see that for high masses ($M_*>10^{10}$ M$_\odot$), the merger timescale is approximately constant with redshift, whereas it decreases somewhat with redshift for lower masses. 

We assume that a single (redshift-independent) power law fit in stellar mass is sufficient to describe the high-mass behaviour (which is the usual regime of interest). In particular, we find that the following fit works well: 
\begin{equation}
T_\mathrm{20}^{500}(M_*,z)=2\hspace{0.5mm}\mathrm{Gyr}\times \bigg(\frac{M_*}{10^{10}\mathrm{M}_\odot} \bigg)^{-0.55},
\label{eq:TSfitsimple}
\end{equation}
with the uncertainty in the normalisation and slope equal to 0.2 Gyr and 0.05, respectively. This approximation is shown by blue lines in both panels of Fig.~\ref{fig:figTS}. We see that this works fairly well in the chosen mass regime. The typical deviation of the true values from the fit is up to $15$ per cent, but only for $M_*>10^{10}$ M$_\odot$. For lower masses ($M_*<10^{10}$ M$_\odot$) this fit becomes progressively worse (with both mass and redshift). 

For mass threshold samples (samples with galaxy masses above $M_*$), we find a similar fit, with normalisation of $1.15$ Gyr and slope $-0.38$. This is applicable for the popular selections $M_*>10^{10}$ M$_\odot$ and $M_*>10^{11}$ M$_\odot$, yielding constant merger timescales of $1.15$ and $0.48$ Gyr, respectively. For the selection $M_*>10^{9.5}$ M$_\odot$, this simple fitting formula overpredicts the merger timescale by up to $30$ per cent at most redshifts, and even more at $z>4$.

From Fig.~\ref{fig:figTS}, we can see that the merger timescale for a given mass is approximately a power law in mass at all redshifts. In order to capture the full behaviour of the merger timescale, for arbitrary mass and redshift, we assume the following formula (as a replacement of Eqn.~\ref{eq:TSfitsimple}) at every redshift:
\begin{equation}
\log_{10}T_\mathrm{mg}=b+a\log_{10}\bigg(\frac{M_*}{10^{10}\,\mathrm{M}_\odot}\bigg),
\label{eq:TSfitM}
\end{equation}
where $b$ is the normalisation and $a$ the slope of the stellar mass dependence. These fits are shown by black lines in the left panel of Fig.~\ref{fig:figTS}. 

We find that the fitting parameters depend significantly on redshift. For $z<7.5$ we find that the following fit for the merger timescale at a given mass works well:
\begin{equation}
T_\mathrm{20}^{500}(M_*,z)=T_0\hspace{0.5mm}\mathrm{e}^{b(z-z_0)^3}\bigg(\frac{M_*}{10^{10}\mathrm{M}_\odot} \bigg)^{a_0+a_1(1+z)^{a_2}},
\label{eq:TSfitfin}
\end{equation}
with the parameters given in Table~\ref{tab:tab1}. The fits given by this formula are shown in the right hand panel of Fig.~\ref{fig:figTS}. We see that they capture the behaviour of the merger timescale reasonably well. This fit deviates no more than $10$ percent for $z>1$, but the error relative to the true values can be as large as $25$ percent at $z=0$, largely due to the local features in our merger timescale. In particular, we find that our simpler fit (\ref{eq:TSfitsimple}) works better for very high masses ($M_*>10^{11}$ M$_\odot$). Finally, we note that we have performed the same fit to mass threshold samples, giving the merger timescale $T_\mathrm{mg}(M_\mathrm{star}>M_*)$. The relevant parameters are also given in Table~\ref{tab:tab1}.

\begin{table*}
\begin{center}
\caption{Parameters of merger timescale fitting formulae. Formulae denoted as $T_\mathrm{20}^{500}$ refer to merger timescales appropriate for close pairs selected with $r_\mathrm{max}=20 $ $h^{-1}$kpc and $v_\mathrm{max}=500$ kms$^{-1}$. \textit{Left}: Merger timescale for arbitrary mass and redshift. For large stellar masses ($M_*>10^{10}$), these formulae can be replaced with a redshift-independent power law in stellar mass, given by Eqn.~(\ref{eq:TSfitsimple}). Fits are given both for single values of $M_*$ and for mass threshold samples, i.e. samples of close pairs chosen such that the primary galaxy has a stellar mass larger than $M_*$. \textit{Right}: The variation of merger timescales with selection criteria. This formula can be used to obtain a merger timescale with an arbitrary selection of $r_\mathrm{max}$ and $v_\mathrm{max}$, or to convert a close pair fraction from one selection to another.
\label{tab:tab1}
}
\end{center}
\begin{tabular}{c|c}

\begin{minipage}{.5\linewidth}
\begin{tabular}{l}
  $T_\mathrm{20}^{500}(M_*,z)=T_0\hspace{0.5mm}\mathrm{e}^{b}\bigg(\frac{M_*}{10^{10}\mathrm{M}_\odot} \bigg)^{a}$, \\
  $b(z)=b_0(z-z_0)^3$, $a(z)=a_0+a_1(1+z)^{a_2}$
\end{tabular}\\
\begin{tabular}{|c|c|c}
  \hline \hline
  \hspace{10mm} & $T_\mathrm{mg}(M_*,z)$ \hspace{5mm} & $T_\mathrm{mg}(M_\mathrm{star}>M_*,z)$\\
  \hline \hline
  $T_0$ [Gyr]\hspace{3mm} & $1.432\pm0.028$  \hspace{5mm} & $1.119\pm0.028$\\
  \hline
  $b_0$\hspace{8mm} & $-0.011\pm0.002$ \hspace{5mm} & $-0.0019\pm0.0006$\\
  \hline
  $z_0$\hspace{8mm} & $3.31\pm0.04$ \hspace{5mm} & $2.98\pm0.36$\\
  \hline
  $a_0$\hspace{8mm} & $-0.601\pm0.0027$ \hspace{5mm} & $-0.521\pm0.041$\\ 
  \hline
  $a_1$\hspace{8mm} & $0.147\pm0.031$ \hspace{5mm} & $0.138\pm0.029$\\
  \hline 
  $a_2$\hspace{8mm} & $0.54\pm0.08$ \hspace{5mm} & $0.58\pm0.11$\\
  \hline 
\end{tabular}
\end{minipage} &

\begin{minipage}{.3\linewidth}
\begin{tabular}{l}
  $T_\mathrm{mg}=T_\mathrm{20}^{500} \bigg(\frac{r_\mathrm{max}}{20\mathrm{h}^{-1}\mathrm{kpc}}\bigg)^\alpha \frac{\mathrm{erf}\big(v_\mathrm{max}/V_0\big)^\beta}{\mathrm{erf}\big(500\hspace{0.3mm}\mathrm{kms}^{-1}/V_0\big)^\beta}$ 
\end{tabular}\\
\begin{tabular}{|c|c|}
  \hline \hline
  $\hspace{5.5mm}V_0$\hspace{7.5mm} & $(540\pm30)$ kms$^{-1}$\hspace{5.5mm} \\
  \hline
  $\hspace{5.5mm}\alpha$\hspace{7.5mm} & $1.32\pm0.1$\hspace{5.5mm} \\
  \hline
  $\hspace{5.5mm}\beta$\hspace{7.5mm} & $0.78\pm0.05$\hspace{5.5mm} \\ 
  \hline 
\end{tabular}
\end{minipage}


\end{tabular}
\end{table*}

\subsection{Dependence on close pair selection criteria}
\label{sec:TSrv}

We now consider how the merger timescale depends on the pair selection criteria $r_\mathrm{max}$ and $v_\mathrm{max}$. Merger timescales inherit these dependencies from the close-pair fraction. We begin by analysing the dependence on $r_\mathrm{max}$. Previous theoretical studies have only attempted to determine this dependence for a few values (e.g. KW08). KW08 find that the dependence is approximately linear, and this assumption has been adopted in other studies (e.g. Xu et~al. \citealt{Xu}).  Observationally, de Ravel et~al.~\cite{deRavel} studied the dependence in detail, finding a steeper slope than KW08 (1.2 vs. 1.0). However, de Ravel et~al. came to this conclusion using their full sample (not split by stellar mass or redshift).

\begin{figure*}
\includegraphics[width=0.95\textwidth, trim = 0 20 0 0]{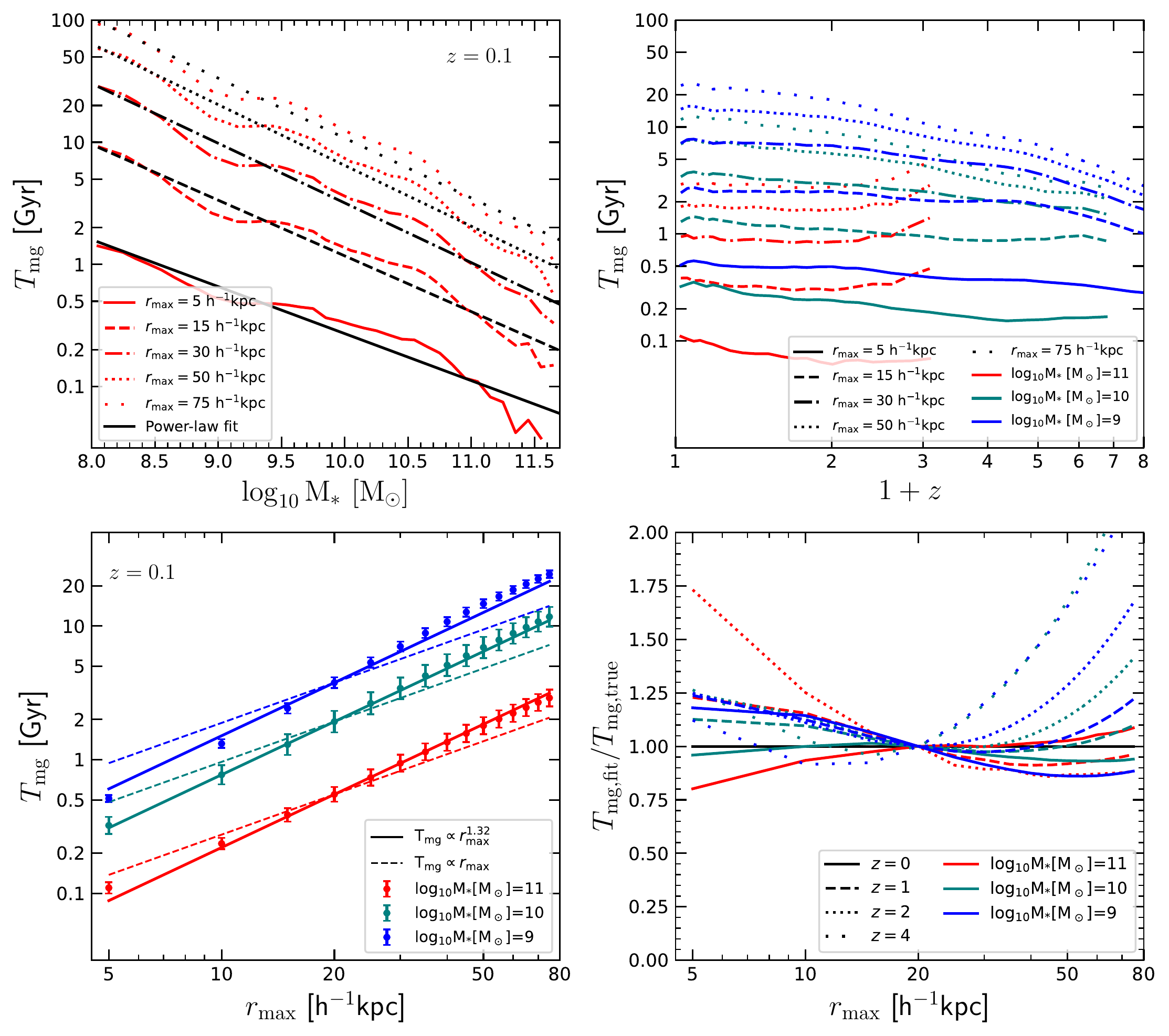}
\caption{Merger timescale for conversion of major ($\mu_* \in [0.25,1]$) close pair fractions to merger rates, for different pair selection criteria $\vert v_\mathrm{sep}\vert<500$ kms$^{-1}$ and $r_\mathrm{sep}<r_\mathrm{max}$, and $r_\mathrm{max}$ between 5 and 75 $h^{-1}$kpc. {\it Top left}: Merger timescale as a function of stellar mass for several close pair selections as given by the legend. Red lines show the results from {\tt GALFORM}, while black lines show power law fits at $z=0.1$, given by Eqn.~(\ref{eq:TSfitM}). {\it Top right}: Merger timescale as a function of redshift for three different masses and several close pair selections, as given by the legend. Colours indicate different masses, while line types represent different close pair selections. {\it Bottom left:} Dependence of merger timescale on maximum projected separation for several masses. Symbols and error bars ($1\sigma$-confidence intervals) represent results from {\tt GALFORM} for 15 selections, while lines show fits. Dashed lines show a linear fit $T_\mathrm{mg}\propto r_\mathrm{max}$ normalised at $r_\mathrm{max}=20$ $h^{-1}$kpc, while solid lines show a fit of the form $T_\mathrm{mg}\propto {r_\mathrm{max}}^{1.32}$. {\it Bottom right:} Ratio of our fitting formula to the model prediction for merger timescales as a function of $r_\mathrm{max}$ for several masses and redshifts.}
\label{fig:figTSr}
\end{figure*}

The top left panel of Fig.~\ref{fig:figTSr} shows the merger timescale as a function of stellar mass at $z=0.1$ for several close pair selections. The mass dependence varies little as the selection is changed, with the main distinction being a change in normalisation. The slope appears constant with mass. The top right panel of Fig.~\ref{fig:figTSr} shows the dependence of the merger timescale on redshift for several close pair selections, as well as different masses. The dependence on redshift remains the same for all selections, as long as merger timescales are viewed for a fixed mass bin.

The bottom left panel of Fig.~\ref{fig:figTSr} shows the dependence of the merger timescale on $r_\mathrm{max}$ for several masses at $z=0.1$. We first consider a linear fit $T_\mathrm{mg}\propto r_\mathrm{max}$ normalised at $r_\mathrm{max}=20$ $h^{-1}$kpc (which is the value for which we studied the merger timescale as a function of stellar mass and redshift in the previous section). This fit, advocated by KW08, underpredicts the merger timescale for large maximum projected separations and underpredicts it at small separations. We therefore adopt an alternative fit $T_\mathrm{mg}\propto {r_\mathrm{max}}^{\alpha}$, and calculate $\alpha$ for different mass bins and redshifts. 

The slope, $\alpha$, varies with  mass and redshift but is generally constrained to be between $1.1$ (for high-mass galaxies) and $1.9$ (for low-mass galaxies). The low-mass slope can be attributed to large numbers of spurious pairs, whose numbers should grow as $\propto r^{2}$. Our high-mass dependence, $T_\mathrm{mg}\propto r_\mathrm{max}^{1.1}$, is closer to the slope of $1.0$ found by KW08.

Despite this variation with mass, we adopt the best-fitting value $\alpha=1.32$. This fit is shown by the solid lines in the bottom left panel of Fig.~\ref{fig:figTSr}. It approximates merger timescales much better than the linear fit at $z=0.1$. In order to explore the validity of the fit at higher redshifts, we plot the ratio $T_\mathrm{mg,\hspace{0.5mm}fit}/T_\mathrm{mg,\hspace{0.5mm}true}$ in the bottom right panel of Fig.~\ref{fig:figTSr} as a function of $r_\mathrm{max}$, for several masses and redshifts. The fit deviates significantly for massive galaxies ($M_*=10^{11}$ M$_\odot$) with close pair selection criteria ($r_\mathrm{max}<10 $ $h^{-1}$kpc), with the deviation increasing with redshift (implying a redshift dependence of the slope $\alpha$). The same is found for intermediate and low mass galaxies ($M_*<10^{10}$ M$_\odot$) at large separations ($r_\mathrm{max}>30$ $h^{-1}$kpc) and high redshifts ($z>4$).

We note that considering a restricted subset of the full range of separations ($r_\mathrm{max}\in[10,30]$ $h^{-1}$kpc out of $r_\mathrm{max}\in[5,75]$ $h^{-1}$kpc) leads to deviations between the fit and our results of no more than $25\%$ at all masses and redshifts. With this choice, our merger timescale fit (Eqn.~\ref{eq:TSfitfin}) becomes:
\begin{equation}
T_\mathrm{mg}(M_*,z,r_\mathrm{max})=T_\mathrm{20}^{500}(M_*,z)\bigg(\frac{r_\mathrm{max}}{20\hspace{0.5mm}h^{-1}\mathrm{kpc}}\bigg)^{1.32\pm0.1},
\label{eq:TSfitfinr}
\end{equation}
where $T_\mathrm{20}^{500}(M_*,z)$ is the merger timescale for the selection limits  $r_\mathrm{max}=20$ $h^{-1}$kpc and $v_\mathrm{max}=500$ kms$^{-1}$, given by Eqn.~(\ref{eq:TSfitfin}) and in Table~\ref{tab:tab1}. While the range in which this formula works very well ($r_\mathrm{max}\in[10,30]$ $h^{-1}$kpc) might be somewhat small, we note that this covers most selections adopted in the literature.

\begin{figure}
\includegraphics[width=\columnwidth, trim = 0 20 0 0]{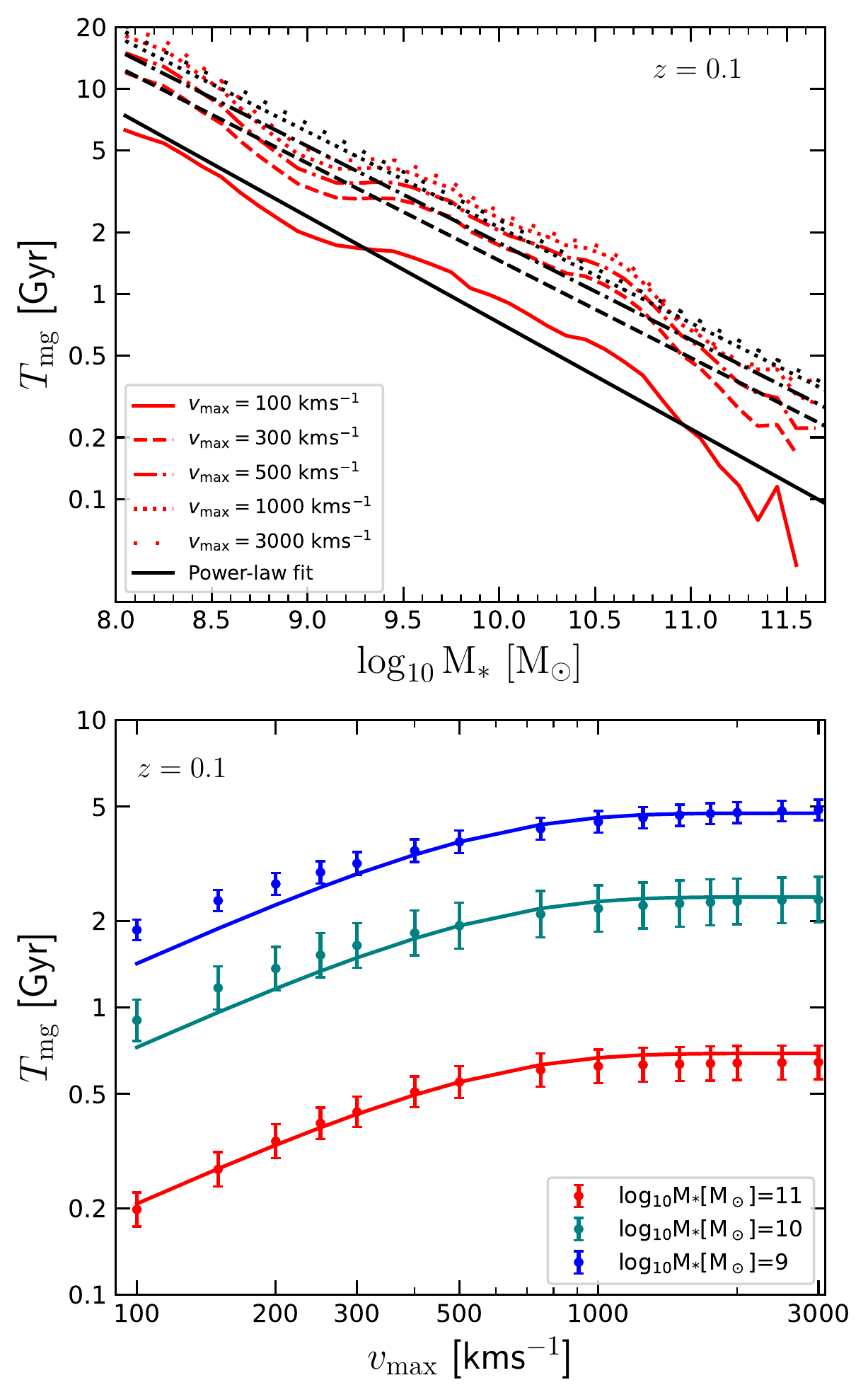}
\caption{Merger timescale for conversion of major ($\mu_* \in [0.25,1]$) close pair fractions to merger rates, for different pair selection criteria $\vert v_\mathrm{sep}\vert<v_\mathrm{max}$, with $v_\mathrm{max}$ varying between 100 and 3000 kms$^{-1}$, and $r_\mathrm{sep}<20$ $h^{-1}$kpc. {\it Top \st{left}}: Merger timescale as a function of stellar mass for several close pair selections as given by the legend. Red lines show the results from {\tt GALFORM}, while black lines show power law fits at $z=0.1$, given by Eqn.~(\ref{eq:TSfitM}).
{\it Bottom \st{left}:} Dependence of merger timescale on maximum relative line-of-sight velocity for several masses. Symbols and error bars ($1\sigma$-confidence intervals) represent results from {\tt GALFORM} for 15 selections, while lines show fits according to Eqn.~(\ref{eq:TSfitv}). 
}
\label{fig:figTSv}
\end{figure}

We now turn to the dependence of the merger timescale on the maximum velocity separation (with maximal separation kept fixed). From the top panel of Fig.~\ref{fig:figTSv} we can see that the dependence on stellar mass is decoupled from the dependence on $v_\mathrm{max}$. Increasing $v_\mathrm{max}$ only results in a rescaling of the relation between merger timescale and mass. Similarly, changing $v_\mathrm{max}$ results in a rescaling of the dependence on redshift. From the top panel of Fig.~\ref{fig:figTSv} it is apparent that the merger timescale (or number of pairs) saturates by some maximal velocity. The bottom panel of Fig.~\ref{fig:figTSv} shows this explicitly. For low values of $v_\mathrm{max}$, the merger timescale is approximately a power law in $v_\mathrm{max}$ (with a best-fitting slope of 0.78) Across all masses and redshifts, we find a saturation by $v_\mathrm{max}=1000$ kms$^{-1}$ which can be encapsulated with the following formula:
\begin{equation}
T_\mathrm{mg}\propto\mathrm{erf}\bigg(\frac{v_\mathrm{max}}{V_0}\bigg)^\beta.
\label{eq:TSfitv}
\end{equation}
In reality, the parameters $V_0$ and $\beta$ depend on stellar mass and redshift, but taking mean values works well. We find best fitting values $V_0=(540\pm30)$ kms$^{-1}$ and $\beta=0.78\pm0.05$.  The fit is shown with solid lines in the bottom \st{left} panel of Fig.~\ref{fig:figTSv}, showing that it works well at $z=0.1$ for a few masses. More generally, we find that the approximate values are within $15\%$ of the true ones as long as $v_\mathrm{max}$ is within $[300,3000]$ kms$^{-1}$, independent of stellar mass and redshift.

\subsection{An approximate formula for the merger timescale}

With the adoption of the fits described in previous subsection, our final merger timescale formula as a function of stellar mass and redshift, as well as pair selection criteria, can be written as:
\begin{equation}
\begin{split}
T_\mathrm{mg}(M_*,z,r_\mathrm{max},v_\mathrm{max})=&T_\mathrm{20}^{500}(M_*,z)\times\bigg(\frac{r_\mathrm{max}}{20\hspace{0.3mm}h^{-1}\mathrm{kpc}} \bigg)^\alpha\\
                                    &\times\frac{\mathrm{erf}(v_\mathrm{max}/V_0)^\beta}{\mathrm{erf}(500\hspace{0.3mm} \mathrm{kms}^{-1}/V_0)^\beta}.
\label{eq:TSfitfinv}
\end{split}
\end{equation}
The parameters of this formula are given in Table~\ref{tab:tab1}. We give separate parameters for close pair samples selected at stellar mass $M_*$, and samples selected with a threshold value $M_*$. We remind the reader that the functional form of Eqn. (\ref{eq:TSfitfinv}) was derived by considering the dependence of the merger timescale on each of the four variables individually (with others kept fixed), but the parameters themselves were not. They were derived by finding the best fit in the 4D space of merger timescale values. A simpler version of the formula, applicable to high-mass galaxies ($M_*>10^{10}$) uses  Eqn.~(\ref{eq:TSfitsimple}) for $T_\mathrm{20}^{500}(M_*,z)$.

The $r_\mathrm{max}-$dependent factor in Eqn.~(\ref{eq:TSfitfinv}) is equal $1$ at $r_\mathrm{max}=20$ $h^{-1}$kpc, while the $v_\mathrm{max}$ dependency is somewhat more complicated. The error function is different from $1$ for all values of its argument, and our default value $v_\mathrm{max}=500$ kms$^{-1}$ is not close to the regime of saturation in our fitting formula. The constant denominator is present to ensure that the $v_\mathrm{max}$ dependency evaluates to $1$ at $v_\mathrm{max}=500$ kms$^{-1}$.

Our fit works best for $r_\mathrm{max}\in[10,30]$ $h^{-1}$kpc and $v_\mathrm{max}>300$ kms$^{-1}$, with the discrepancy relative to the true values typically less than $15$ per cent (at worst 25 per cent, depending on mass and redshift). Regardless of the possible error, we argue that it is better to apply our formula than to use merger timescales which are not appropriate to the sample selection for a measured close pair fraction. This is because the pair fraction depends strongly on $r_\mathrm{max}$ and $v_\mathrm{max}$, so ignoring these dependencies can lead to significant discrepancies. Equivalently, when comparing different close pair fraction results, it is better if these are converted to a standard selection using our scaling relations.

In order to validate our merger timescale formula directly, we apply it to close pair samples that we measure from the simulation. This results in an inferred merger rate that in principle should be equal to the one measured directly from the simulation, and it should not depend on the selection. The close pair samples we choose for this comparison are intended to represent realistic selections: we use $r_\mathrm{max}\in[5,20]$ $h^{-1}$kpc and $r_\mathrm{max}\in[10,30]$ $h^{-1}$kpc (two popular choices, see Table~\ref{tab:tab0}), as well as a wide selection of $r_\mathrm{max}\in[5,50]$ $h^{-1}$kpc. We combine these selections with $v_\mathrm{max}<300$ kms$^{-1}$ and $v_\mathrm{max}<500$ kms$^{-1}$ (again, two popular choices), as well as $v_\mathrm{max}<3000$ kms$^{-1}$ (by which value the number of pairs has saturated; this selection matches pair fractions calculated through photometric redshift differences). In total, this gives nine close pair selections.

\begin{figure*}
\includegraphics[width=\textwidth, trim = 0 20 0 0]{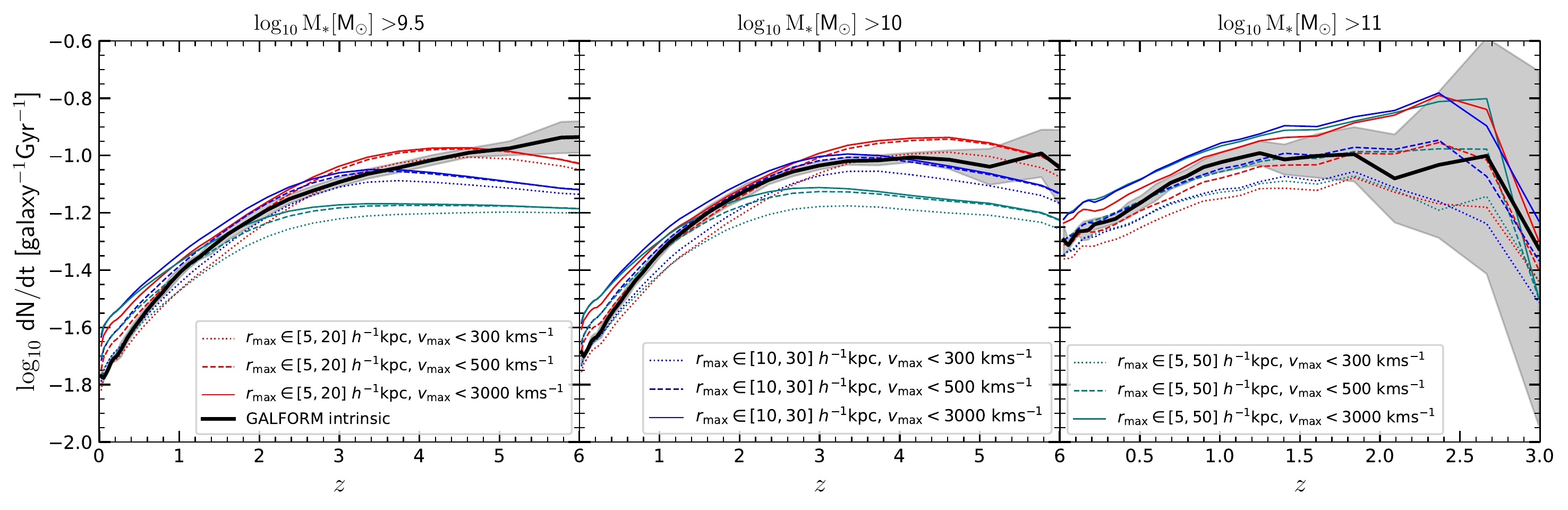}
\caption{Comparison of merger rates calculated directly from the simulation (black lines and shadings) with those inferred from various close pair selections (coloured lines, as per the legends). Each panel represents a different mass selection as given by the titles. The inferred merger rates are obtained by dividing close pair fractions measured in the simulation by the merger time-scale given by Eqn.~(\ref{eq:TSfitfinv}). }
\label{fig:figMergRateDirectInferred}
\end{figure*}

In Fig.~\ref{fig:figMergRateDirectInferred} we compare the dependence on redshift of the true and inferred merger rates for these nine selections. These comparisons are shown for the three popular mass selections previously used in this work. The inferred merger rates for $r_\mathrm{max}\in[5,20]$ $h^{-1}$kpc show the best agreement with the true one; this is not surprising since our fit was centred on a similar selection ($r_\mathrm{max}\in[0,20]$ $h^{-1}$kpc). Merger rates inferred from $r_\mathrm{max}\in[10,30]$ $h^{-1}$kpc pair samples show similar levels of agreement. For $r_\mathrm{max}\in[5,50]$ $h^{-1}$kpc, the inferred merger rate underestimates the real one beyond $z=3$ for the two lower mass selections, by up to a factor of two. However, this selection is not usually used by observational works. Overall, we find that our formula for the merger time-scale does not reproduce the merger rate perfectly, but it represents a significant improvement over previously available ones (e.g. Kitzbichler \& White \citealt{KitzWhite}).

\subsection{Comparison with other models for merger timescale}

It is worth comparing our predicted merger timescales to previously published ones. Some studies assume an additional factor $C_\mathrm{mg}$ in the relation between merger rates, close pair fractions and merger timescales (Eqn.~\ref{eq:eqCPTS}), which represents the probability of merging. The merger rate is then given by $\mathrm{d}N/\mathrm{d}t=C_\mathrm{mg}\times f_\mathrm{maj}/T_\mathrm{mg}$. However, the probability of merging $C_\mathrm{mg}$ is often taken to be constant (e.g. Lotz et~al. \citealt{Lotz2011}). We compare our merger timescale with $T_\mathrm{mg}/C_\mathrm{mg}$ for studies which take $C_\mathrm{mg}\neq1$, since we  assume $C_\mathrm{mg}=1$. 

\begin{figure*}
\includegraphics[width=\textwidth, trim = 0 20 0 0]{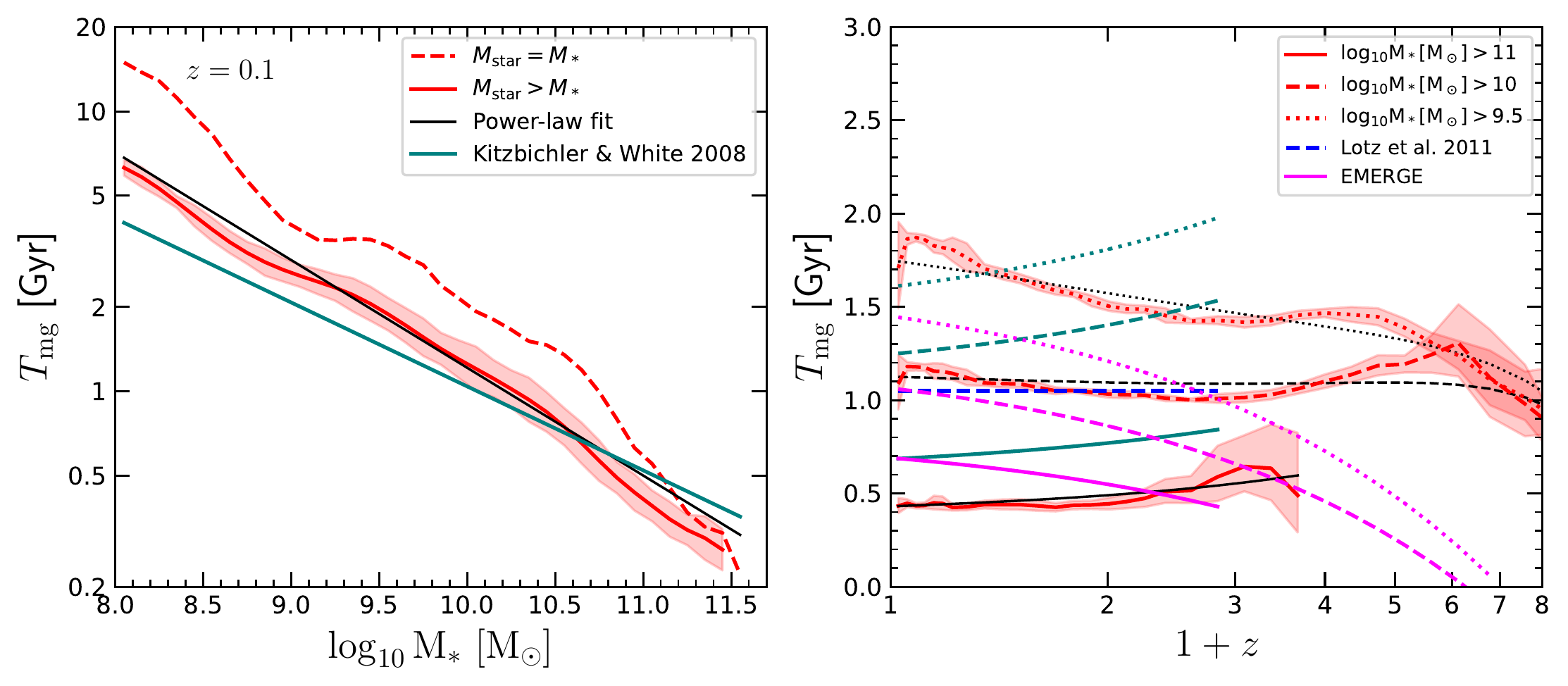}
\caption{Merger timescale for conversion of major ($\mu_* \in [0.25,1]$) close pair fractions to merger rates, with pair selection criteria $r_\mathrm{sep}<20 $ $h^{-1}$kpc and $\vert v_\mathrm{sep}\vert<500$ kms$^{-1}$. Lines are discontinued at masses or redshifts where the merger rate drops to zero (giving an infinite merger timescale) due to no merging galaxies being detected. Red lines represent results from {\tt GALFORM}, while black lines represent the fitting function given by Eqn.~(\ref{eq:TSfitfin}), and parametrised in Table~\ref{tab:tab1}. Other coloured lines are from other works as per the legend. {\tt EMERGE} results are from O'Leary et~al.\protect\cite{OLeary}. Shaded regions represent $1\sigma$-confidence intervals. {\it Left}: Merger timescale for galaxies of (dashed line) and above (solid line) stellar mass $M_*$, at $z=0.1$. {\it Right}: Merger timescale for three threshold mass selections (given by the legend) as a function of redshift. 
}
\label{fig:figTSComparison}
\end{figure*}

KW08 studied the merger timescale as a function of stellar mass and redshift, but their formulae are for galaxies with stellar masses \textit{above} a threshold mass $M_*$. The left panel of Fig.~\ref{fig:figTSComparison} shows our merger timescale for galaxies of mass $M_*$ and above $M_*$, along with results from KW08 (in particular, we compare our merger timescale with their formula, Eqn.~(\ref{eq:KW}), rescaled to our standard selection: $r_\mathrm{max}=20$ $h^{-1}$kpc and $v_\mathrm{max}=500$ kms$^{-1}$). Our mass threshold values, $T_\mathrm{mg}(M_\mathrm{star}>M_*)$ are always lower than values at a given mass, $T_\mathrm{mg}(M_\mathrm{star}=M_*)$, since $T_\mathrm{mg}$ almost monotonically decreases with stellar mass.  We find that our merger timescale exhibits a somewhat steeper dependence on mass than KW08: at $z=0.1$ we find that a power law fit (black line in left panel of Fig.~\ref{fig:figTSComparison}) has a slope $a=-0.38$, while KW08 report $a=-0.3$. Furthermore, unlike KW08 we find that the slope, $a$, changes with redshift.

The right panel of Fig.~\ref{fig:figTSComparison} shows the dependence of our merger timescale on redshift, alongside a fit to this dependence (black lines), and predictions from other models. The KW08 timescale is $50\%$ larger than ours for galaxies with $M_*>10^{11}$ M$_\odot$, at all redshifts. At $z=0$ we find that the merger timescales agree for $M_*>10^{10}$ M$_\odot$, and ours is higher for $M_*>10^{9.5}$ M$_\odot$. KW08 found $T_\mathrm{mg}\propto(1+z/8)$ for all masses. This is consistent with our results only for $M_*>10^{11}$ M$_\odot$. We also compare our results with the {\tt EMERGE} semi-empirical model (O'Leary et~al. \citealt{OLeary}). {\tt EMERGE} predicts a falling merger time-scale for all mass selections. For $M_*>10^{11}$ M$_\odot$ the two merger timescales roughly agree, although we predict no fall with redshift. For other mass selections, our merger timescale is generally higher in normalisation and falls less quickly with redshift. Finally, we make a comparison for stellar masses $M_*>10^{10}$ M$_\odot$ with the results of hydrodynamical simulations from Lotz et~al.~\cite{Lotz2011}. These results agree very well with ours, with both implying a merger timescale of 1 Gyr and very little evolution in redshift.

Overall, we find that different models can predict very different merger timescales. Some models might agree in their predictions of merger rates, while disagreeing in their close pair fractions, which leads to disagreeing timescales. These disagreements are apparent in both their value (normalisation) and redshift evolution. We note, however, that all results shown in Fig.~\ref{fig:figTSComparison} agree, at least qualitatively, that the redshift evolution is weak at most. These conclusions are very different to those obtained from the Illustris simulation by Snyder et~al.~\cite{Snyder}, who find $T_\mathrm{mg}\propto(1+z)^{-2}$. We note that this strong redshift dependence is the result of rising merger rates (Fig.~\ref{fig:figMRRObs108} and \ref{fig:figMRRObsLar}) and fairly constant close pair fractions, which are too low compared to observations (Fig.~\ref{fig:figCPObsR}).

A direct observational test of the accuracy of different merger timescale predictions is not possible.
It might be argued that the KW08 timescale is superior since they constructed light cones to measure their close pair fraction. However, as we have shown in \S~\ref{sec:CPF}, our close pair fractions are in good agreement with observed ones, at least to the level of agreement between different observational studies (Fig.~\ref{fig:figCPObsR}). Furthermore, for unresolved subhaloes, the model in KW08 uses the Lacey \& Cole\cite{LaceyCole} formula for subhalo merger times, while {\tt GALFORM} uses a more accurate formula (Simha \& Cole et~al. \citealt{Simha}, Eqn.~\ref{eq:TmergSimha}). Our approach uses the full simulation volume, which means that we are able to include all mergers in our calculation; this is not the case with a lightcone. 


Finally, in addition to the dependence on stellar mass and redshift, we can compare our predictions to observational measurements of pair fractions as a function of selection criteria. Our average dependence on maximal separation ($T_\mathrm{mg}\propto r_\mathrm{max}^{1.32}$) is similar to that implied by the observational study of close pairs by de Ravel et~al.~\cite{deRavel}, who find $\alpha=1.24$. This small disagreement is expected since their study includes only bright galaxies, which inherently have a higher proportion of physical pairs (driving the fit towards smaller values of $\alpha$; we find $\alpha=1.1$ for massive galaxies, as discussed in \S~\ref{sec:TSrv}). As the authors note, these results are comparable to the observed projected two-point galaxy correlation function $w_\mathrm{p}(r_\mathrm{p})$. This is because the correlation function represents the excess probability of finding a galaxy pair at distance $r$ relative to a uniform distribution. However, care needs to be taken in the comparison, since correlation functions always remove the contribution from spurious pairs (while the close pair fraction includes them). The projected two-point correlation function is often assumed to be a power law, $w_\mathrm{p}\propto r_\mathrm{p}^{\gamma+1}$, where $\gamma$ is the slope of the 3D two-point correlation function, $\xi(r_\mathrm{p})$. This leads to $f_\mathrm{pair}\propto r_\mathrm{p}^{3+\gamma}$, at least in regimes where we expect pairs to be physically associated (i.e. high-mass systems, see Fig.\ref{fig:figCPMR}). Galaxy clustering measurements from the SDSS (Li et~al. \citealt{Li}, Zehavi et~al. \citealt{Zehavi}) and GAMA (Farrow et~al. \citealt{Farrow}) both found $\gamma=-1.8$, implying $\alpha=1.2$. Our high-mass slopes (up to 1.1) are consistent with these findings. Le Fèvre et~al. \cite{LeFevre} find $\gamma=-1.7$ ($\alpha=1.3$) in VIMOS, in even better agreement with our results.




\section{Summary and conclusion}
\label{sec:sum}

We have used an updated version of the {\tt GALFORM} semi-analytical galaxy formation model, with more accurate tracking of subhalo orbits, to study galaxy merger rates, close pair fractions and merger timescales with unprecedented precision. This is possible due to the large volume of the Planck Millennium simulation, as well as the large number of outputs. We are able to probe merger statistics with high precision in mass (40 bins in stellar mass between $10^8$ and $10^{12}$ M$_\odot$) and redshift (40 redshift bins between $z=0$ and $z=10$). 

Our results can be summarized as follows:
\begin{itemize}
\item We predict a rapid decrease in the major merger rate per galaxy and close pair fraction at high stellar mass ($>M_*\approx 10^{11.3}$ M$_\odot$ at $z=0$), in agreement with recent observations. This drop is due to the exponential suppression of galaxy abundance seen in the galaxy stellar mass function (GSMF). The stellar mass at which this drop occurs reduces to $M_*\approx 10^{10.5}$ M$_\odot$ by $z=4$, again following the behaviour of the GSMF. This drop also causes merger-related quantities at fixed stellar mass $M_*$ to decline at some redshift $z$. This is the redshift at which $M_*$ galaxies enter the exponentially suppressed regime in the GSMF.
\item The stellar mass dependence of the major merger rate predicted by {\tt GALFORM} agrees well with observations and the Illustris simulation at $z=0$. The merger rate per galaxy evolves to reach a maximum before declining above some mass-dependent redshift; this agrees with most observations, but disagrees with the Illustris and EAGLE hydrodynamical simulations, as well as semi-empirical models, which predict a merger rate that continues to increase with redshift. This turnover is possibly a result of the GSMF in {\tt GALFORM} decreasing rapidly with redshift for massive galaxies, whereas observational data suggests that the GSMF declines more weakly with redshift in this regime.
\item We have performed an extensive comparison of our predicted close pair fraction with observations and other theoretical models. In agreement with most results, as a function of redshift our close pair fraction shows a maximum and then a decline, depending on the stellar mass selection. The details of this behaviour are not well constrained by observations, nor do models converge on a unified picture. We have provided precise predictions for close pair fractions up to very high redshifts ($z=10$) to help build a unified picture of galaxy clustering and merging.
\item The close pair fraction and corresponding merger timescale depend on maximum projected separation as $\propto r_\mathrm{max}^\alpha$, with the slope $\alpha$ decreasing from values close to 2 at low masses, to values close to 1 at high masses. This behaviour is due to low-mass galaxies predominantly having projected pairs, while high-mass galaxies mostly have physical pairs. Despite the variation with stellar mass and redshift, we find that $\alpha=1.32$ works well as an approximation in the range $r_\mathrm{max}\in[10,30]$ $h^{-1}$kpc. This slope is in agreement with observational studies of the small-scale clustering of galaxies, but it differs somewhat from previous findings that suggest a linear dependence. We find that the close pair fraction depends on maximum velocity separation as $f_\mathrm{maj}\propto v_\mathrm{max}^{0.78}$ for low values and saturates by $v_\mathrm{max}=1000$ kms$^{-1}$ for all masses and redshifts. 
\item We provide a formula for the average major merger timescale of close pairs which works well for all masses and redshifts, as well as close pair selection criteria $r_\mathrm{max}$ and $v_\mathrm{max}$:
\begin{equation}
\begin{split}
T_\mathrm{mg}(M_*,z,r_\mathrm{max},v_\mathrm{max})=&T_\mathrm{20}^{500}(M_*,z)\times \bigg(\frac{r_\mathrm{max}}{20\hspace{0.3mm}h^{-1}\mathrm{kpc}} \bigg)^{1.32}\\
&\times\frac{\mathrm{erf}(v_\mathrm{max}/V_0)^{0.78}}{\mathrm{erf}(500\hspace{0.3mm}\mathrm{kms}^{-1}/V_0)^{0.78}},
\label{eq:TSfitfinrv}
\end{split}
\end{equation}
where $V_0=540$ kms$^{-1}$. This formula works best for $r_\mathrm{max}\in[10,30]$ $h^{-1}$kpc and $v_\mathrm{max}>300$ kms$^{-1}$, but can also be extrapolated outside of these regimes. The error function can be expanded out at velocities not close to saturation ($v_\mathrm{max}<500$ kms$^{-1}$), giving $T_\mathrm{mg}\propto v_\mathrm{max}^{0.78}$. 

\item Our merger timescale selected with  $r_\mathrm{max}=20$ $h^{-1}$kpc and $v_\mathrm{max}=500$ kms$^{-1}$, $T_\mathrm{20}^{500}(M_*,z)$, can be well approximated as a redshift-dependent power law in stellar mass. The fitting function is given by Eqn.~(\ref{eq:TSfitfinv}), with relevant parameters given in Table~\ref{tab:tab1}.  We find that the merger timescale for massive galaxies ($M_*>10^{10}$ M$_\odot$) is approximately redshift-independent, and is well described by
\begin{equation}
T_\mathrm{20}^{500}(M_*,z)=2\hspace{0.5mm}\mathrm{Gyr}\times \bigg(\frac{M_*}{10^{10}\mathrm{M}_\odot} \bigg)^{-0.55}.
\label{eq:TSfitsimple2}
\end{equation}
For close pair samples chosen with masses {\it above} a threshold value $M_*$, a similar formula can be used, but with a normalisation of $1.15$ Gyr and slope $-0.38$.

\end{itemize}

Our focus in this work has been on the statistics of mergers, as mergers are an important process in galaxy formation. Upcoming synoptic surveys and high-redshift observations will be able to test our predictions on close pair fractions in fine detail. In a future paper we will investigate the importance of mergers vs. star formation in the buildup of the stellar mass of galaxies. We will look at the contributions of different merger types to this growth. Furthermore, the role of mergers in the growth of spheroids will be compared with disc instabilities, alongside star formation in bursts caused by both mechanisms.




\section*{Acknowledgements}

F. H. would like to acknowledge support from  the Royal Astronomical Society, the European Union's Erasmus+ programme and the Science Technology Facilities Council through a CDT studentship (ST/P006744/1). This work was also supported by STFC grant ST/T000244/1. F.H. would like to thank Joseph O'Leary for providing data from the {\tt EMERGE} model. We thank the reviewer for their comments, which helped make this a better work. This work used the DiRAC@Durham facility managed by the Institute for Computational Cosmology on behalf of the STFC DiRAC HPC Facility (www.dirac.ac.uk). The equipment was funded by BEIS capital funding via STFC capital grants ST/K00042X/1, ST/P002293/1, ST/R002371/1 and ST/S002502/1, Durham University and STFC operations grant ST/R000832/1. DiRAC is part of the National e-Infrastructure.

\section*{Data availability}

The data underlying this article will be provided upon request to the corresponding author.

\bsp	
\label{lastpage}
\end{document}